\documentclass[runningheads]{llncs}

\usepackage[T1]{fontenc}
\usepackage{lmodern} 
\usepackage[paperwidth=155mm,paperheight=235mm,textwidth=12.2cm,textheight=19.3cm,headsep=16pt,includehead,centering]{geometry}
\usepackage[figuresright]{rotating} 
\makeatletter
\newenvironment{hsidewaysfigure}[1]{%
  \edef\@tempa{\noexpand\@float{figure}[\fps@figure]}\@tempa
  \refstepcounter{figure}\label{#1}\addtocounter{figure}{-1}%
  \begin{lrbox}\rot@float@box
  \begin{minipage}\textheight
}{\end@rotfloat}
\makeatother

\usepackage{graphicx}
\usepackage{amsmath}
\usepackage{xcolor} 

\usepackage{hyperref}
\hypersetup{colorlinks=true,linkcolor=blue,citecolor=blue,urlcolor=blue,
  hypertexnames=false, 
  bookmarksdepth=2,bookmarksnumbered=true,bookmarksopen=true,bookmarksopenlevel=1}
\newcommand{\citet}[1]{\cite{#1}}
\newcommand{\citep}[1]{\cite{#1}}

\title{When AI Tutors Speak: Evidence from a Randomized Field Experiment  }
\titlerunning{When AI Tutors Speak}

\author{Shihao Yang\thanks{Corresponding author.} \and Marshall Van Alstyne \and Chrysanthos Dellarocas}
\authorrunning{S. Yang et al.}

\institute{Department of Information Systems, Questrom School of Business, Boston University, Boston, MA, USA\\
\email{\{syang536,mva,dell\}@bu.edu}\\[8pt]
\normalfont\itshape Working draft --- version of September 20, 2026}

\begin{document}
\maketitle
\vspace{-11mm}
\begin{abstract}
Students increasingly study alongside generative artificial intelligence (AI), yet unguided access to fluent answers invites cognitive offloading, and there is little evidence on which configurations of AI tutoring produce learning. Two design margins are usually bundled together: pedagogical structure (how the tutor teaches) and interaction modality (how students talk to it). We separate them. In a preregistered randomized field experiment in a graduate corporate-finance course of an online MBA, we randomized 86 students between a structured tutor grounded in the course materials and a holdout in which consumer AI remained freely available. Within the tutored arm, each student's channel alternated weekly between voice and text, so the modality effect is identified within student. Structure mattered: tutored students gained 6.63 points more than ability-matched peers ($p=.007$), and the gain was concentrated in written reasoning, where the share of answers reaching relational quality rose from 8\% to 49\% in the tutored arm against 8\% to 27\% in the holdout. Modality did not matter for learning. The instructor's own final, on file for all 86 randomized students, shows the same direction (2.6 points of 100, with no difference on a pre-treatment midterm). Voice nearly doubled conversational interaction and cost 2.8$\times$ as much to deliver, yet it produced weekly mastery statistically equivalent to text, even as students came to prefer it. Pedagogical structure shapes what students practice, and modality shapes how they interact with the tutor. Making an AI more humanlike does not by itself make it more educational.
\end{abstract}

\section{Introduction}
One-on-one tutoring is among the most effective interventions in education and among the least scalable. Online graduate programs feel this constraint acutely. They serve working professionals who study asynchronously, between job and family obligations: the students for whom an always-available personal tutor would matter most, in the setting where human tutoring is least feasible. Large language models have made conversational tutoring deployable at program scale. Since students will study with AI in any case, the open question is which configurations of AI tutoring produce authentic learning.

How students already use AI sharpens that question. Unguided access to a fluent answer engine invites cognitive offloading (the delegation of thinking to an external aid \citep{risko2016offloading,sparrow2011google,fisher2015searching}), and the classroom evidence is now causal. In a field experiment in high-school mathematics, unrestricted GPT-4 access improved assisted practice but significantly harmed subsequent unassisted exam performance, while a tutor-configured version with pedagogical guardrails eliminated the harm \citep{bastani2025}. At far larger scale, 30 months of records on more than 26,000 Chinese secondary students show that generative-AI adoption raised homework scores while lowering examined performance \citep{stromberg2026penalty}. Other studies converge on the same pattern: AI assistance that lifts performance while active leaves little behind once removed \citep{darvishi2024agency,gerlich2025ai,lee2025critical,kosmyna2025brain}, and consultants who used ChatGPT to complete data-science tasks far beyond their training retained no more technical knowledge than peers who worked unaided, which led the authors to call the tool an exoskeleton \citep{wiles2024exoskeleton}. Students will not stop using AI. The design question is therefore whether an AI built as a tutor, one that asks before it answers and makes the student produce rather than consume, converts AI use from offloading into practice. This study uses AI to counter the offloading that AI invites.

Within that agenda, this paper examines the configuration choice about which current evidence says the least: modality. Essentially all field evidence on LLM tutoring concerns typed chat, yet voice AI now sustains natural spoken dialogue, and learning science gives concrete reasons why speaking with a tutor should differ from typing at one. Answering a tutor's question is retrieval practice rather than passive review \citep{roediger2006test,karpicke2011retrieval,mcdermott2021retrieval}; producing the answer aloud adds the distinctive encoding of the production effect \citep{macleod2010production,ozubko2012production,macleodbodner2017production}; spoken dialogue elicits the self-explanations that drive learning \citep{chi1994selfexplanations,chi2009aci}; and voice is among the strongest cues that an artificial agent has a humanlike mind, which invites the social engagement that sustains use \citep{schroeder2016mistaking,epley2007seeing}. Whether these mechanisms survive contact with a real course (real deadlines, multitasking, exam pressure) is an empirical question that no field experiment has yet answered.

We answer it with a preregistered randomized field experiment\footnote{Pre-analysis plan, preregistered on OSF before the field period: \url{https://osf.io/yr8c9}.} embedded in a live module of Boston University Questrom's Online MBA (MX720, Module~2, Weeks~8--12; $N=86$ consented, with $N=51$ completing both assessment waves; the completed five-week field period is reported end to end). The design has two randomized layers. Between students, we randomize consenting participants to a structured conversational AI tutor grounded in the course materials or to a no-AI holdout. Within the AI condition, each student's tutor alternates weekly between voice and text, with the starting modality randomized. Every multi-week participant therefore experienced both modalities on new course content, so the voice--text contrast is identified within student, free of selection into modality. Because the setting is quantitative finance, both modes allowed students to request tutorials with visual display of graphs and equations (rendered in \LaTeX) on a persistent session blackboard. The platform also instruments the tutoring process itself (turn-level timing, dialogue density, question-asking, engaged time), so we can trace modality effects to the learning behaviors that produce them and relate engagement dose to gains. The instructor's own midterm, weekly checks, and final are also on file for every randomized student and for the 395 classmates who never enrolled, which lets us check the main result on an instrument we did not design and on a sample that did not attrit.

The paper makes four contributions. \emph{First}, it provides causal field evidence that a pedagogically structured tutor improves learning over the realistic counterfactual of self-study with ambient consumer AI. Under the full preregistered specification, AI-arm students gained 6.6 more points (of 55) than ability-matched holdout peers ($p=.007$, robust across two scoring panels). The gain was concentrated in the conceptual depth of written answers, where the share reaching relational quality grew sixfold, rather than in ceilinged multiple-choice scores. The instructor's own final, on the intact randomization of 86 students, shows the same direction (2.6 points of 100, $p=.040$), and his pre-treatment midterm shows no difference between arms. \emph{Second}, it provides, to our knowledge, the first causal within-student field evidence on voice versus text AI tutoring in a live graduate course. Week-level mastery is statistically equivalent between modalities, while the interaction itself changes substantially (roughly twice the dialogue turns per minute and 2.4$\times$ the question-asking on equal engaged time). Voice is therefore better understood as an engagement and adoption lever than as a per-minute learning advantage. \emph{Third}, it shows that ambient consumer AI did not close the structured-tutor gap. Holdout students used consumer AI freely throughout the study (21--42\% in any given week), yet the tutor's randomized advantage holds over that realistic environment, and removing ambient-AI users from the control group leaves the estimate essentially unchanged. \emph{Fourth}, it provides conversation-level cost accounting from provider billing that prices this configuration choice. Voice's process transformation carries a 2.8$\times$ cost premium, so programs can buy engagement with voice, or equivalent measured weekly mastery with text at a third of the cost.

Two measurement frameworks recur throughout the paper, so we fix them at the outset. \emph{Bloom's taxonomy} \citep{anderson2001taxonomy} orders the cognitive demand of a task, from remembering and understanding through applying and analyzing to evaluating and creating. We use it both to construct assessment items at graded difficulty and to drive the tutor's questioning upward through those levels. The \emph{SOLO taxonomy} \citep{biggs1982solo} grades the quality of an open-ended \emph{answer} by its structure rather than its correctness: a single relevant point (unistructural), several unconnected points (multistructural), points integrated into a coherent argument (relational), or a principle generalized beyond the prompt (extended abstract). Bloom thus describes what a question demands, and SOLO describes what an answer delivers. Our depth outcomes count answers that reach the relational level or better (Table~\ref{tab:taxonomies}; measurement details in Section~3.7).

\begin{table}[!htbp]
\centering
\caption{The two taxonomies as used in this study.}
\label{tab:taxonomies}
{\small
\setlength{\tabcolsep}{4pt}
\begin{tabular}{lll}
\hline
Level & Bloom (item difficulty) & SOLO (answer quality; score) \\
\hline
lower & remember / understand & prestructural (0); unistructural (1) \\
      & apply                 & multistructural (2) \\
higher & analyze / evaluate    & relational (3) \\
      & create                & extended abstract (4) \\
\hline
\end{tabular}}
\end{table}

\subsection{Research Questions and Hypotheses}
This study evaluates whether a structured conversational AI tutor improves learning in an authentic graduate course, and in particular whether delivery modality (the axis this design is specifically built to identify) matters for learning.

\begin{enumerate}
    \item \textbf{RQ1 (AI access).} Does access to the AI tutor increase learning gain (pre$\rightarrow$post) relative to a no-AI holdout? \textbf{H1 (primary):} students randomized to AI access exhibit larger gains in total content score than holdout students.
    \item \textbf{RQ2 (modality).} Among students with AI access, does tutoring modality (voice vs.\ text) affect learning? \textbf{H2 (primary):} within the AI group, mastery is higher in weeks delivered via voice than via text.
    \item \textbf{RQ3 (cognitive demand).} Does the access effect differ by item cognitive demand? \textbf{H3 (secondary):} the effect is larger for higher-order Bloom items (analyze\slash evaluate\slash create) than for lower-order items (understand\slash apply).
    \item \textbf{RQ4 (dose-response).} Among students with AI access, does learning gain scale with engagement intensity? \textbf{H4 (secondary):} gains increase with tutor engagement intensity.
    \item \textbf{RQ5 (depth).} Does AI access improve deeper learning as measured by SOLO-scored short answers? \textbf{H5 (secondary):} the effect on SOLO depth outcomes is at least as large as on MCQ surface performance (its two SOLO outcomes are preregistered co-primary endpoints, reported uncorrected alongside H1).
\end{enumerate}

\section{Related Work}
The study rests on a three-link argument set against a cautionary backdrop. The backdrop is that unguided AI use promotes cognitive offloading, and that pedagogical structure separates AI that harms learning from AI that helps. The three links are that engagement is the binding constraint of educational technology; that voice makes an AI tutor a more humanlike, and therefore more engaging, interlocutor; and that socially engaged interaction is the kind that produces learning. Each link maps onto a preregistered hypothesis.

\subsection{AI tutoring and the offloading risk}
Two bodies of field evidence frame the intervention. On one side, structured AI tutoring works. AI instructional agents increase perceived learner control and learning efficiency relative to self-paced platforms \citep{qin2025ai}, and a randomized crossover experiment in a large undergraduate physics course found that students learned more in less time with a structured AI tutor than in an active-learning class session, with higher engagement and motivation \citep{kestin2025}. On the other side, unguided generative AI can undermine the learning it appears to support. The sharpest evidence manipulates exactly this contrast: unrestricted GPT-4 access harmed unassisted exam performance while a guardrailed tutor configuration of the same model did not \citep{bastani2025}. AI assistance on a learning platform improved output quality while active but left no gains once withdrawn \citep{darvishi2024agency}, a reliance pattern also found among knowledge workers, whose confidence in the tool predicts less enacted critical thinking \citep{lee2025critical}. A workplace experiment separates the two outcomes cleanly: consultants given ChatGPT for data-science tasks outside their skill set scored 18 to 49 percentage points higher than untreated peers, close to the level of professional data scientists on two of three tasks, yet afterwards they answered technical questions without the tool no better than the control group. The authors call the tool an exoskeleton, since it extends what a worker can do without teaching the worker to do it \citep{wiles2024exoskeleton}. At much larger scale the pattern is the same. Across 30 months of data on more than 26,000 Chinese secondary students, generative-AI adoption raised homework scores by 18\% and cut completion time by 30\% but lowered monthly exam performance by 20\% and entrance-exam scores by 18\% and 24\%, with the losses concentrated among the roughly 80\% of users whose behavior is consistent with outsourcing homework to the tool \citep{stromberg2026penalty}. Unguided generative AI can thus raise homework performance and speed while lowering examined learning, which is the pattern that attempt-first, mastery-paced tutoring is designed to guard against. The intervention studied here sits deliberately on the guardrailed side of this divide: a tutor that diagnoses, probes, and makes the student produce, deployed in a cohort where ambient consumer AI is freely available (and measured weekly in the holdout arm). These findings motivate the access hypothesis (H1). The design space these deployments draw on is by now well articulated. Mollick et al.\ \citep{mollick2024agents} show that assigning generative models distinct pedagogical roles (mentor, role-player, and instructor-facing evaluator) turns a chat interface into scalable simulated practice, and they are explicit that such systems remain prototypes awaiting rigorous testing. This paper addresses that gap between an articulated design space and field evidence. Closest to our own setting, Bapna et al.\ \citep{bapna2025agentic} deploy a no-code agentic analytics tool, in effect a virtual data scientist addressed in natural language, in a core analytics course of a global MBA program, and pair student--agent interaction logs with closed-book final examination scores to ask whether agent use raises learning and how usage varies across students. That study establishes the graduate management classroom as a site where conversational AI can be evaluated against consequential, instructor-administered assessment, and it shares our reliance on interaction telemetry as the process measure. Our design differs in the two respects that motivate this paper: assignment to tutor access is randomized against a holdout rather than observed, and the delivery channel is itself randomized within student. Every tutoring deployment in this literature, including that one, reaches the student through typed chat. Whether the delivery channel itself matters is untested in the field, and that question is this study's distinctive contribution.

\subsection{Engagement as the binding constraint}
The central failure mode of educational technology is disengagement rather than weak content: completion rates in online courses average 3 to 6 percent \citep{reich2019mooc}. Interventions that move engagement therefore move outcomes at scale. Agrawal et al.\ \citep{agrawal2026personalization} show this causally on a learning platform. In an RCT with 7,750 students, personalized recommendations raised section engagement by 60 percent and platform-wide engagement by 14 percent, and the engagement gains persisted and grew under scaled deployment. Engagement also means more than exposure. In a comparison spanning more than 27,000 MOOC learners, the estimated learning benefit of interactive doing was roughly six times that of watching or reading \citep{koedinger2015doer}. Most directly for conversational agents, Xu et al.\ \citep{xu2022dialogue} find that dialogic reading with a voice-based agent improved children's story comprehension through a measured mediation path of increased verbal engagement. Engagement is therefore the mediator to instrument, which is what our dose-response design does (H4).

\subsection{Voice as a social-engagement technology}
Why should voice raise engagement? The media-equation and Computers-Are-Social-Actors programs established that people mindlessly apply human social rules to media that exhibit human cues, voice chief among them \citep{nass1994computers,reeves1996media,nass2000machines,nass2005wired}. Schroeder and Epley \citep{schroeder2016mistaking} provide the sharpest causal version: adding a voice to machine-generated text makes observers more likely to mistake the machine for a human, while stripping speech down to text makes even humans seem machine-like. Speech likewise conveys thoughtfulness and intellect that identical written content does not \citep{schroeder2015sound}, and a survey of two decades of agent-voice research concludes that voice plays an essentially social role in human-agent interaction \citep{seaborn2021voice}. Anthropomorphism theory supplies the bridge to behavior: humanlike agents satisfy and invite sociality motivation \citep{epley2007seeing}. Consistent with the theory, studies of commercial voice assistants find that perceived humanness and social presence drive engagement and continued use \citep{moriuchi2019okay,moriuchi2021anthropomorphism,mclean2019alexa,fernandes2021voice,wagner2019human}, and an early classroom study of a GPT-4o voice tutor observed students spontaneously treating the tutor socially during programming lessons \citep{jacobs2025voice}.

Social engagement is, finally, the kind of engagement that teaches. Under social agency theory, a human voice and a conversational style prime learners to treat instruction as a conversation, which deepens processing and improves transfer \citep{moreno2000engaging,mayer2003social,atkinson2005voice}. Human-like embodiment helps most when paired with a human voice \citep{mayer2012embodiment}, and contingent spoken dialogue with a media character improves children's science learning through verbal engagement \citep{xu2022contingent}. This chain, from voice to perceived humanness to social engagement to learning, generates our modality hypothesis (H2) and raises its stakes: a voice advantage in the field would show that the chain operates in real coursework.

\subsection{Cognitive production mechanisms}
These strands compose into one testable causal chain rather than an omnibus argument. When an identical tutor moves from text to voice, what changes first is the production cost and the social framing of a student turn: speaking is faster than typing, and speech frames the exchange as conversation rather than composition \citep{schroeder2016mistaking,reeves1996media}. The first observable consequence should therefore be behavioral: more frequent turns, more questions, and more spontaneous thinking aloud on the same engaged time. The learning-relevant step is the last link. Each additional produced turn is retrieval practice \citep{roediger2006test,karpicke2011retrieval}, carries the distinctive encoding of production \citep{macleod2010production,ozubko2012production}, and is often a self-explanation, the highest-leverage activity in the active--constructive--interactive hierarchy \citep{chi1994selfexplanations,chi2009aci}. The chain therefore predicts, in order, a process transformation under voice that is observable in telemetry, and downstream of it a learning difference only if the added production is of the kind that teaches. Our design measures the links separately, and Section~4 reports them separately: the process link holds with near-unanimous within-student consistency, while the outcome link is where the equivalence result disciplines the theory. Section~3.9 fixes the vocabulary this chain requires and distinguishes usage, engaged time, interaction intensity, and attention rather than treating them as one ``engagement'' construct.

\subsection{Modality--task fit and depth measurement}
Task--Technology Fit theory qualifies the prediction. Text persists on screen, so re-reading and side-by-side comparison reduce working-memory load in stepwise analytical work, while voice is sequential and ephemeral \citep{chen2024voice}. For a quantitative module such as the one studied here, these affordances plausibly favor text. This is a moderating claim that we note but do not test, and it is one reason not to expect the setting to be biased toward voice. On the outcome side, surface mastery and conceptual depth can diverge. Short-answer items scored on the SOLO taxonomy capture the structure of reasoning rather than fluency, and a pre-specified rubric keeps depth outcomes comparable across administrations (H5).

\section{Methodology}

\subsection{Research Setting}
The experiment is embedded in a required quantitative-finance module (MX720, Module~2) of Boston University Questrom's Online MBA, a fully asynchronous, cohort-based degree program for working professionals who fit coursework around full-time jobs. During the five study weeks (course Weeks~8--12) the module moved from pricing and customer analytics into the core of corporate finance: capital budgeting, financing choice, and cost of capital. This is quantitative, problem-set-driven material in which students compute and interpret net present values, discount rates, and costs of capital and defend the resulting decisions. Each week follows the program's standard rhythm. Students work through recorded lectures and readings on their own schedule, practice on problem sets, and converge on graded module deliverables due after Week~12. Support comes from discussion boards, periodic live sessions, and instructor office hours. That help is real, but it is scheduled and shared, so most of an asynchronous learner's study hours pass without access to on-demand explanation.

The tutor entered this environment as an optional weekly practice layer on that week's material, available around the clock through a standalone web application. A student signs in under verified identity, and the tutor runs a practice session on the week's learning objectives, grounded only in the module's own lectures, readings, and problem sets (Section~3.3). It replaced no course component and carried no grade, and it was barred from graded-assessment content. The role it fills is the gap that holdout students' open responses would later name (Section~4.4): on-demand explanation and practice between office hours. The study is conducted under Boston University IRB Protocol \#8506E (approved 6/19/2026). We preregistered the design, hypotheses, outcomes, and analysis plan on OSF before the field period (\url{https://osf.io/yr8c9}); the field period ran 7/1/2026 through 8/5/2026, and the post-test window closed 8/15/2026.

Incentives were identical across arms. Students earned \$4 per completed study week (a completed tutoring session in the AI arm; a brief check-in survey in holdout), rising to a \$20 completion payment at four completed weeks plus a \$5 bonus for five-for-five; these weekly payments locked in regardless of the post-test. Completing the post-test also entered students into a raffle of \$600, \$400, and \$200 prizes and earned a free e-book authored by program faculty. All $N=86$ participants completed the pre-test; $N=51$ completed the post-test.

\subsection{Design}
The study has two randomized layers.

\noindent\textbf{Layer 1 (between-student AI access).} We randomize consented students at the student level to one of two groups:
\begin{itemize}
    \item \textbf{AI group:} access to the structured conversational AI tutor each week (plus standard course resources).
    \item \textbf{Holdout group:} standard course resources and a brief weekly survey only; no access to the experimental AI tutor during the study window.
\end{itemize}

\noindent\textbf{Layer 2 (within-student rolling modality; AI group only).} For students in the AI group, the tutor alternates weekly between voice and text. Each student's initial modality (Week~8) is randomized, and modality then alternates across Weeks~8--12. From Week~10 the rotation is completion-based: a missed week's modality carries forward to the next attempt, which keeps modality exposure balanced under partial attrition.

Of the $N=51$ students who completed both the pre- and post-tests, 14 were AI-arm students who started in voice, 18 were AI-arm students who started in text, and 19 were in the no-AI holdout.

\subsection{The tutoring platform}
\textbf{The information system.} The tutor is a purpose-built web application; Figure~\ref{fig:platform} shows the architecture. Sessions run on a cascaded voice pipeline (speech-to-text, LLM, text-to-speech) streamed turn by turn. The tutor brain is GPT-5.4. We chose it on the three criteria the design makes binding: it must follow a long, rule-dense pedagogical prompt reliably (the one-question and attempt-first rules of Section~3.4 must hold on every turn); its streaming latency must be low enough for natural spoken turn-taking; and its structured tool-calling must be dependable, because the tutor writes to the blackboard on nearly every turn. The same model, version-pinned, served both modalities for the entire field period, so the voice--text contrast isolates the delivery channel rather than the model. The LLM is grounded in a course-specific knowledge base (weekly readings, lecture notes, problem sets) and follows a structured tutoring prompt that diagnoses before it explains and pushes the student to produce: answer aloud, work the example, check the result. Alongside the conversation, the tutor keeps a persistent ``blackboard'' of structured notes (definitions, worked examples, rendered formulas, diagrams), written through tool calls; students can copy or export the blackboard for later study. Figure~\ref{fig:ui} shows the session view as students experience it, with each interface element annotated; Appendix~\ref{app:interfaces} shows the student progress page and the instructor monitoring dashboard. The platform enforces the weekly modality assignment at session start and for the duration of the session. The assignment is a hard interface lock rather than an instruction. In voice weeks the typed input box is disabled and the composer reads ``Audio only\,---\,please respond by voice,'' so the student can answer only by speaking; in text weeks the microphone is disabled and the session is typed. Students cannot mix channels within a session or opt out of the assigned one, so the modality assigned for a week is also the channel the student actually used. The platform also injects session context at connect (verified identity, study week, session clock, and cross-session memory of prior topics) and meters engaged time with an idle clamp: credit accrues only within five minutes of the student's last input, with server-side caps, and gates the 20-minute weekly goal. Apart from the channel, the same agent runs in both conditions; pedagogy, knowledge base, and prompts are held constant so the modality contrast isolates delivery. Every session persists a full transcript with per-turn timestamps, blackboard content, and interaction events; from Week~11, typed turns also log think/compose timing and attention paradata, and voice sessions store the full call audio.

\begin{hsidewaysfigure}{fig:platform} 
  \centering
  \includegraphics[width=\textwidth]{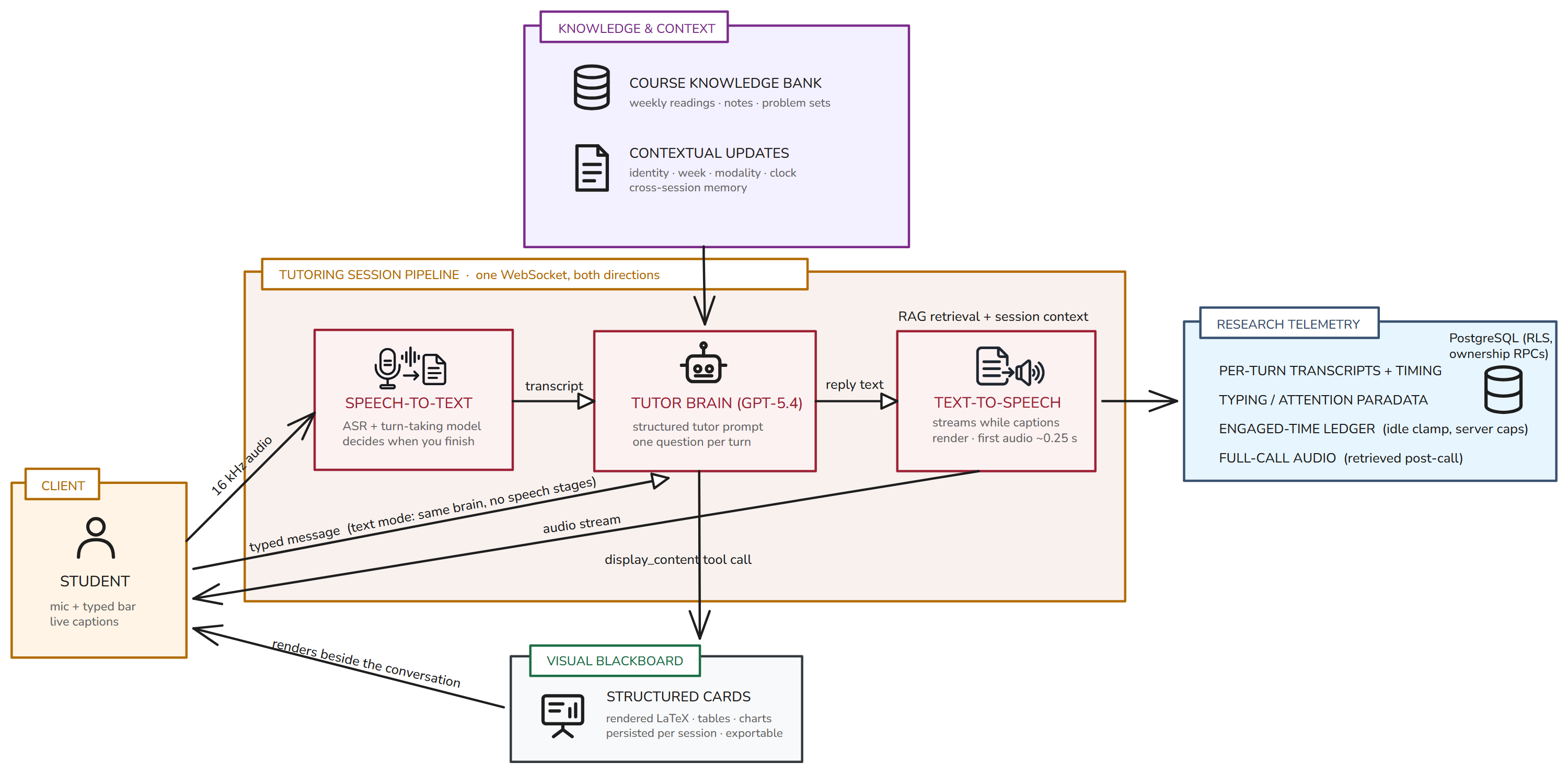}
  \caption{Tutoring platform architecture. Spoken turns travel a cascaded pipeline (speech-to-text, LLM tutor grounded in the course knowledge bank, text-to-speech) over a single WebSocket; typed turns bypass the speech stages and enter the identical tutor brain, which keeps the voice--text comparison clean. Tool calls write the visual blackboard, and every turn persists transcripts, timing, and paradata to the access-controlled research store.}
\end{hsidewaysfigure}

\begin{hsidewaysfigure}{fig:ui}
  \centering
  \includegraphics[width=.78\textwidth,page=1]{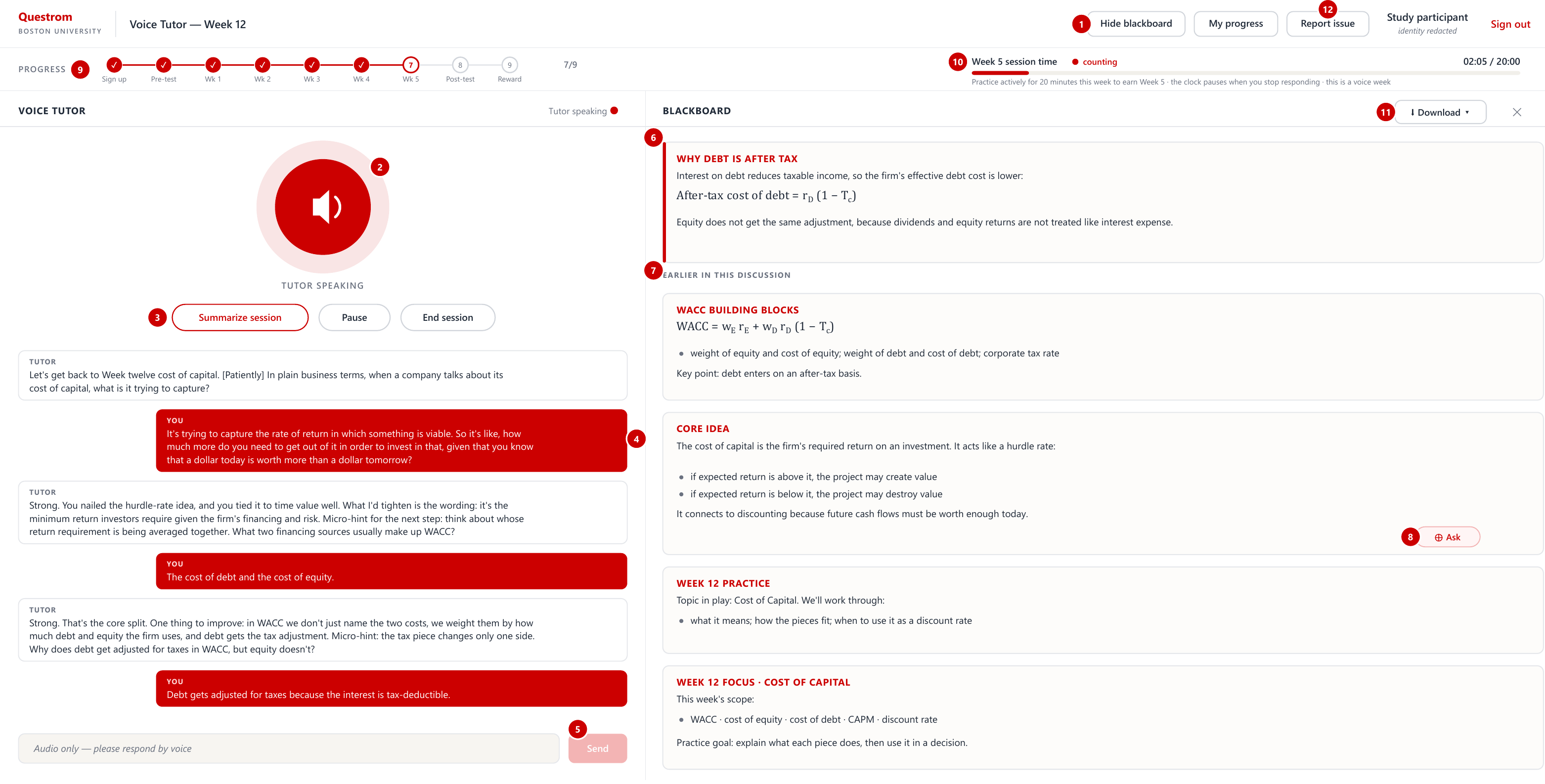}
  \caption{The tutor session view in voice mode, redrawn from the deployed interface for one archived study session (Week~12, cost of capital, 5~August 2026); transcript turns and blackboard cards are reproduced verbatim from the research store, with the student's identity removed. Callouts added. (1)~Header with the blackboard toggle and signed-in identity. (2)~Voice-status orb showing speaking/listening state. (3)~Session controls. (4)~Live transcript; tutor turns carry affect tags. The visible exchange shows the pedagogy of Section~3.4 in operation: the tutor requires an attempt, names what was right, supplies a smaller next step rather than the answer, and closes only when the student states the tax-shield mechanism unaided. (5)~Modality lock: voice weeks disable the typed channel (``Audio only''); text weeks disable the microphone instead. (6)~Blackboard card written this turn, rendered from \LaTeX{}, which consolidates the student's own answer. (7)~Earlier cards persist through the session. (8)~Per-card \emph{Ask} control. (9)~Milestone tracker: sign-up, pre-test, weeks 1--5, post-test, reward. (10)~Weekly session-time counter against the 20-minute engaged-time goal; the clock pauses when the student stops responding (the idle clamp of Section~3.9). (11)~\emph{Download} menu exporting the blackboard and transcript (Markdown, Word, PDF). (12)~\emph{My progress} opens the student dashboard (milestones, weekly tutoring time, and understanding level); \emph{Report issue} files an in-app support report.}
\end{hsidewaysfigure}

\subsection{The tutoring pedagogy}
The treatment is the pedagogy rather than the model; Figure~\ref{fig:cycle} shows the session cycle the prompt enforces (the full prompt is abridged in Appendix~\ref{app:prompt}). A session begins with connection context injected at the socket handshake (verified identity, study week, session clock, and cross-session memory of prior topics), so the tutor greets the student by name, names the week's topic, and opens with a brief \emph{retrieval warm-up}: one or two recall questions on prior material before any new content. The warm-up relies on the testing effect instead of re-presenting notes. The tutor draws its questions from the week's learning objectives in the course knowledge base and orders them up Bloom's ladder (remember/understand $\rightarrow$ apply $\rightarrow$ analyze $\rightarrow$ evaluate; Table~\ref{tab:taxonomies}). It poses \emph{one question at a time} and requires an attempt before it will explain; this is the prompt's central rule. It selects the next question one rung above the level the student has just demonstrated, or laterally if the attempt failed. When a student does not know an answer, the tutor never simply supplies it. It decomposes the problem into a smaller step, offers a graduated hint, or works one example on the shared blackboard and then poses an isomorphic variant back to the student; a nudge protocol re-engages a stalled or disengaged student at an easier entry point. Errors receive targeted feedback in a fixed format (what was right, what was off, one step to repair), and the consolidated result is written to the blackboard, which the student can export. A topic is \emph{complete} only when the student answers a transfer question (the same principle in a novel scenario) without scaffolding; the session then advances to the next objective.

Each element maps onto a hypothesized mechanism. Attempt-first questioning is retrieval practice (H1: guided practice should beat ambient AI's answer-on-demand). The Bloom ladder with a transfer capstone pushes answers from listing toward integration, which is the SOLO relational threshold that the depth outcomes score (H5). Requiring spoken or typed \emph{production} on every turn engages the production and self-explanation channels that motivate the modality contrast (H2), and because the cycle operates identically in both channels, any modality difference isolates delivery rather than pedagogy. Weekly repetition of the cycle is the dose that H4 relates to gains. Among plausible configurations (answer-on-demand assistants, lecture-style explainers, fixed problem sets with AI feedback), this is the one the offloading evidence \citep{bastani2025} identifies as protective, and it operationalizes the interactive tier of the ICAP hierarchy \citep{chi2009aci}: every element forces the student to produce rather than consume.

\begin{figure}[!htbp]
  \centering
  \includegraphics[width=.85\textwidth,page=1]{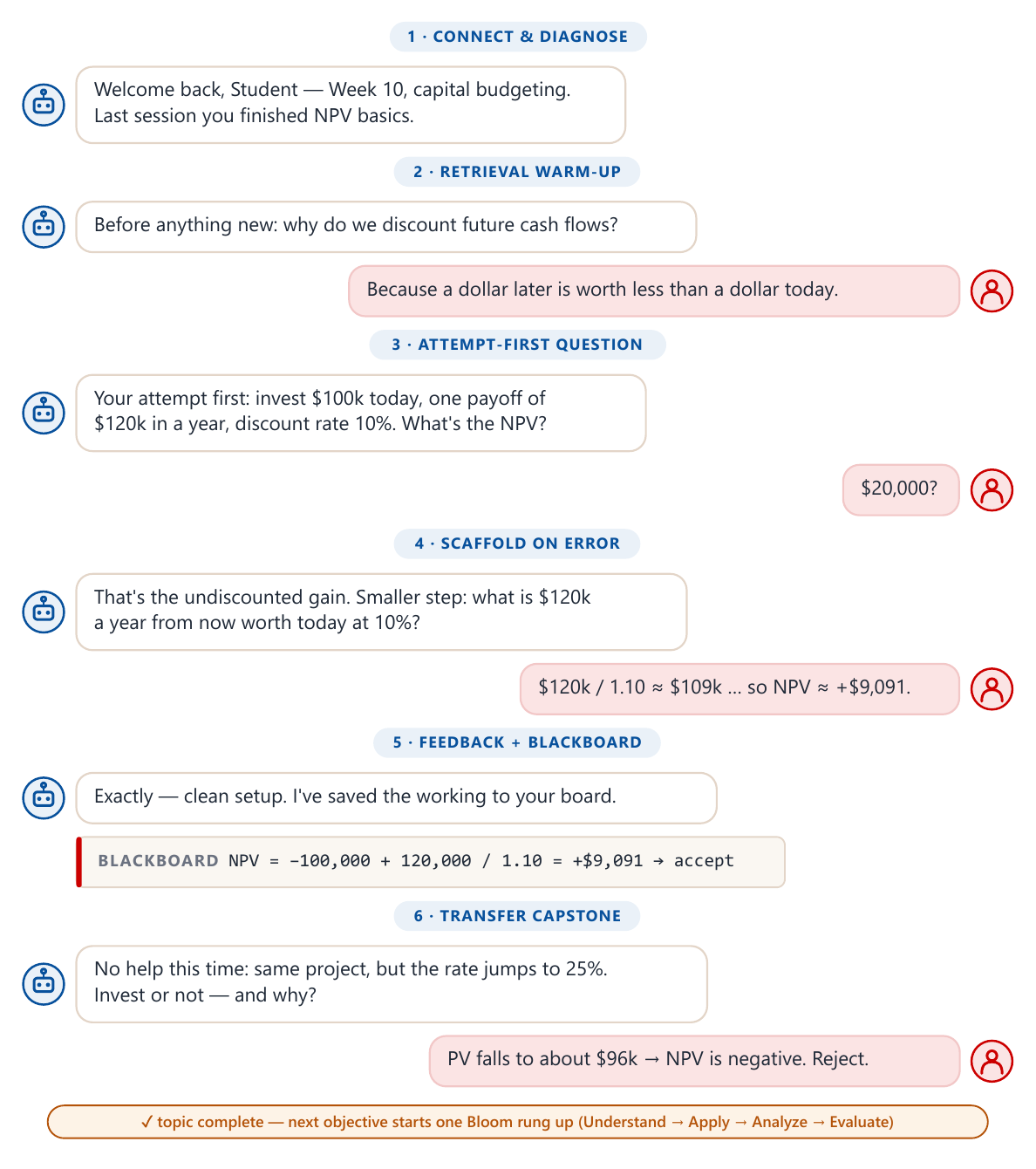}
  \caption{The tutoring cycle enforced by the system prompt, illustrated as one condensed exchange in the platform's session view. The tutor opens with identity and prior-topic memory, warms up with retrieval, requires an attempt before explaining, scaffolds the error with a smaller step rather than the answer, consolidates the worked result to the shared blackboard, and closes the topic only on unassisted transfer; mastery advances the Bloom rung.}
  \label{fig:cycle}
\end{figure}

\subsection{Randomization and realized assignment}
\begin{figure}[!htbp]
  \centering
  \includegraphics[width=.8\textwidth,page=1]{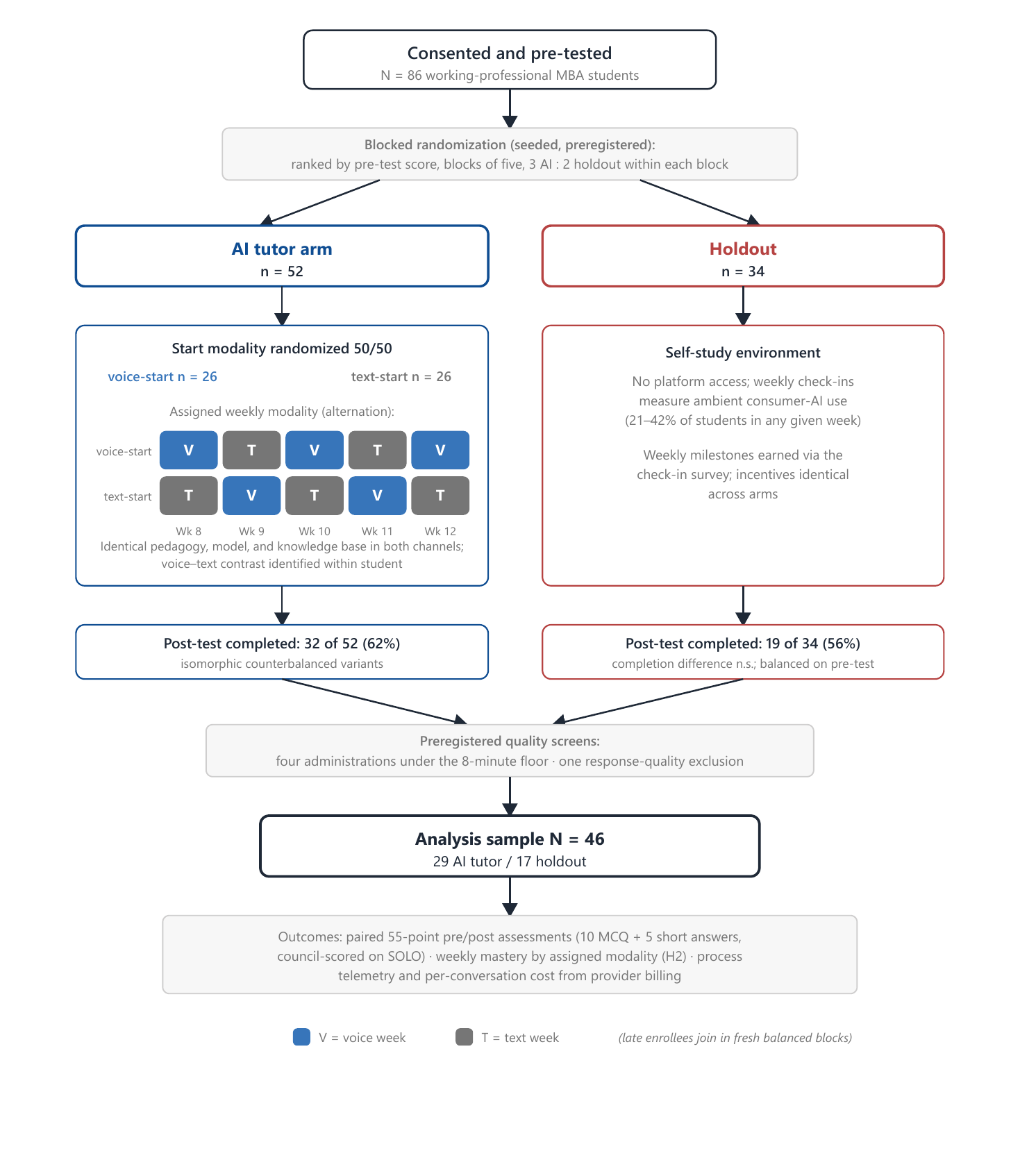}
  \caption{Study design and participant flow. Between students, seeded blocked randomization (stratified on pre-test score) assigns 60/40 to tutor access; within the AI arm, assigned modality alternates weekly from a randomized start, so every multi-week student practices in both channels and the voice--text contrast is identified within student. Preregistered quality screens yield the analysis sample of $N=46$.}
  \label{fig:design}
\end{figure}

\begin{figure}[t]
  \centering
  \includegraphics[width=\textwidth]{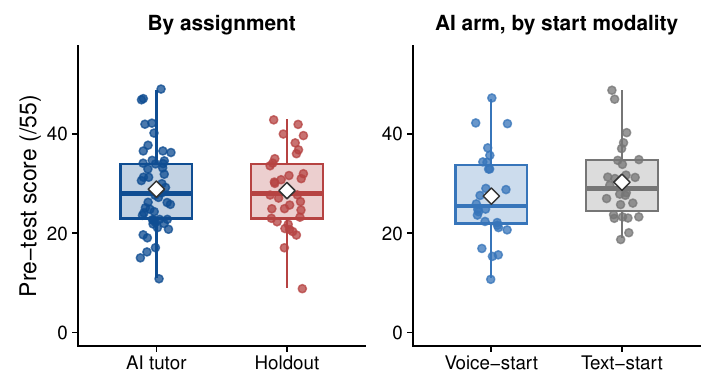}
  \caption{Pre-test balance in the randomized cohort ($N=86$): score distributions are balanced across assignment (52 AI vs.\ 34 holdout) and, within the AI arm, across randomized initial modality (26 voice-start vs.\ 26 text-start). Boxes: interquartile range and median; diamonds: means (AI 28.9 vs.\ holdout 28.6; voice-start 27.5 vs.\ text-start 30.3); the late enrollee, randomized into a fresh block, is included in the means with the pre-test placed on the workbook's 55-point scale (24/55).}
  \label{fig:pretest-distribution}
\end{figure}

\textbf{Arm and start-modality assignment.} We generated the randomization programmatically from a fixed seed recorded with the preregistration, using blocked randomization stratified on the pre-test total score. The routine ranks eligible students (those with a finished pre-test and a valid score) by pre-test total, breaks ties by response identifier, and partitions them into consecutive blocks of five. Within each block it assigns three students to the AI arm and two to the holdout at random; the final partial block is assigned in the same 60/40 proportion. Because adjacent-ranked students land in the same block, the arms are balanced on baseline ability by construction across the full score distribution, not only in expectation. The same seeded generator then randomizes start modality among AI-arm students in a balanced 50/50 draw. The randomized sample comprises $N=86$ consented participants: 52 in AI and 34 in holdout, with the AI group split evenly between voice-start ($n=26$) and text-start ($n=26$). The seeded workbook holds 85 of them; one student enrolled after the seeded batch and was assigned by the same routine into a fresh block. Figure~\ref{fig:design} summarizes the participant flow through both randomized layers, the weekly alternation schedule, and the path to the analysis sample; Figure~\ref{fig:pretest-distribution} shows the realized baseline balance.

\noindent\textbf{Item-variant counterbalancing.} Every assessment item exists in two isomorphic variants, A and B (Figure~\ref{fig:variants}). Each student sees one variant of each item at pre-test and is \emph{guaranteed} the other variant at post-test, so no student ever meets the same question twice, while across the cohort each variant appears at both administrations about equally often. Any difficulty difference between variants therefore cancels out of within-student gains by construction rather than by assumption. In implementation, each participant receives a 15-bit assignment vector from a seeded deterministic generator (bit $k$ selects the pre-test variant of item $k$), and the post-test always serves the exact complement of that vector. ``Derived at read time and never stored'' refers only to the complement: because the post-test form is computed as the bitwise complement of the stored pre-test vector, no separate post-test assignment exists to drift out of sync, and question reuse is impossible by construction. The generator draws vectors under a balanced scheme (each item's variant exposure split as close to 50/50 as possible cohort-wide, then shuffled and dealt), keys them to opaque study codes that carry no identifying information into the survey platform, and runs idempotently, so late enrollees join in fresh balanced batches. Appendix~\ref{app:assignment} lists the core routine.

\subsection{Timeline}
Students take a paired assessment twice: a pre-test before the Week~8 content is released and an isomorphic post-test after the Week~12 deliverables are due. The intervention runs across Weeks~8--12.

\subsection{Measures and scoring}
\textbf{Primary outcome.} The learning outcome is the post-test total content score (55 points), with the pre-test score as a baseline covariate.

\noindent\textbf{Item construction.} Each administration contains 15 items spanning Weeks~8--12 (three items per week, mapped to that week's learning objectives): 10 multiple-choice questions (MCQ) and 5 short-answer items. Item generation followed a blueprint-first pipeline. An assessment blueprint fixed, for each of the 15 slots, the week, the construct drawn from that week's stated learning objectives, the format, the Bloom tag, and a difficulty target. We then drafted candidate items with LLM assistance, grounded in the module's own instructional materials (readings, lecture notes, problem sets), and curated every candidate against the blueprint: editing stems for construct fidelity, checking distractor plausibility, and reworking any item whose key was ambiguous. The LLM was a drafting aid; the blueprint and human curation were the authorities of record. Every item carries a worked solution key, and we validated the MCQ keys by reproducing the survey platform's independent auto-scoring exactly on every administration. Every item was written as a pair of isomorphic variants that share the same deep structure, solution path, and difficulty target and differ in surface parameters and cover context, so the post-test assesses identical skills on unseen surfaces; MCQ correct-answer letters differ between variants by design, so answer-pattern memory cannot transfer. Comparability of the variant pairs is protected twice over: by construction (same solution path, changed numbers and cover story) and by the counterbalanced design above, which cancels any residual variant-difficulty asymmetry out of within-student gains. Items are tagged by Bloom level (lower-order: Understand\slash Apply; higher-order: Analyze\slash Evaluate\slash Create) to support the H3 interaction, and the five short-answer items target explanatory depth rather than recall. Timing behavior from the pre-test administration (median 18 minutes against a 25-minute target) was the calibration check on length and difficulty.

\noindent\textbf{Taxonomies.} Two standard frameworks organize item difficulty and answer quality (Table~\ref{tab:taxonomies}). \emph{Bloom's taxonomy} \citep{anderson2001taxonomy} orders cognitive demand from remembering and understanding through applying, analyzing, evaluating, and creating; we used it to construct the assessment blueprint (each week: two MCQs at apply/analyze and one short answer at evaluate/create) and to drive the tutor's question laddering. In H3, \emph{higher-order} denotes analyze/evaluate/create items (9 of 15) and \emph{lower-order} denotes understand/apply (6 of 15). The \emph{SOLO taxonomy} \citep{biggs1982solo} scores the \emph{structure} of an open response rather than its correctness: prestructural (no valid engagement), unistructural (one relevant point), multistructural (several relevant but unconnected points), relational (points integrated into a coherent argument), and extended abstract (the principle generalized to a novel situation). The masked council described below scores short answers to these five levels. SOLO $\geq 3$ (relational or better) is the preregistered threshold for meaningful conceptual understanding, and the depth outcomes are the mean SOLO level and the share of a student's answers at or above that threshold.

\noindent\textbf{MCQ scoring.} Each MCQ is scored correct/incorrect and weighted to 3 points.

\noindent\textbf{Short-answer scoring (SOLO).} Short-answer items are scored on the pre-specified five-level SOLO rubric (prestructural~$=0$ to extended abstract~$=4$; Table~\ref{tab:taxonomies}), with levels converted to item points by a fixed mapping ($0,1,2.5,4,5$; maximum 5 points per item, 25 across the five items), blind to arm, wave, and identity. The mapping was frozen with the rubric before any outcome was observed, and it is deliberately non-linear: no credit floor for prestructural answers, exactly half credit at multistructural, and its two largest steps at the transitions the construct turns on (accumulating substance, then integrating it), so that assembling an argument is worth more than naming another fact, and the 80\% mark coincides with the preregistered SOLO~$\geq 3$ depth threshold. Because graders return raw levels and both depth outcomes are computed on raw levels, only the total score passes through the mapping, and the primary estimate is insensitive to it: under an equal-spaced 0--5 mapping H1 is $+6.02$ ($p=.009$), and on the source key's native scale (MCQ scored /10, short answer /20 at one point per SOLO level) it is $+3.68$ of 30 points ($p=.007$). The standardized effect ($d=+1.05$) and the share of scale (12.3\% versus 12.1\%) are the same under every weighting. For blinding and handling, responses are keyed by opaque study code and shuffled across students and administrations, pre-scored anchor examples calibrate each item block, and a share of responses repeats under different codes as a within-grader reliability probe. An LLM council drawn from four independent providers does the scoring: GPT-5 (OpenAI; \texttt{openai/gpt-5}), Gemini~2.5~Pro (Google; \texttt{google/gemini-2.5-pro}), Claude Opus~5 (Anthropic; \texttt{claude-opus-5}), and DeepSeek-V3 (\texttt{deepseek/deepseek-chat}). Each model grades every response independently against the rubric. The item score is the council's lower median (even-panel splits round down, per the plan); responses whose four scores span two or more SOLO levels are flagged for human adjudication (16\% of responses), and cross-provider agreement is reported as linearly weighted kappa (mean $0.81$, range $0.76$--$0.84$). A same-provider four-model panel (Claude Fable~5, Opus~5, Sonnet~5, and Haiku~4.5), scored in parallel, is the scoring-sensitivity check (mean $\kappa=0.82$; all effect magnitudes agree), and a human-scored cross-validation subsample accompanies the journal version. Depth outcomes include the change in mean SOLO score and the change in the share of short-answer items at SOLO $\geq 3$.

\subsection{Weekly surveys}
Both arms complete brief weekly self-reports. AI-arm students answer a post-session survey each week covering modality preference on a five-point scale (voice only to text only), session quality by modality, confidence in the week's material, trust in the tutor's answers, and study-habit changes; the repeated preference item supports a within-student preference-trajectory analysis that accompanies H2. Holdout students complete a weekly check-in covering their learning experience and their use of consumer AI tools (ChatGPT, Gemini, Claude, and similar) for coursework that week, including which tools. The holdout's self-reported AI use enters the analysis in two preregistered ways. First, it is a manipulation check that characterizes the realized control condition, which is self-study \emph{with} ambient consumer AI rather than an AI-free environment. Second, it is a measured contamination covariate for interpreting H1, whose estimand is accordingly the tutor's marginal value over a realistic self-study environment in which consumer AI remains freely available. Weekly usage rates in the holdout arm also bound the scope for crossover-style dilution when interpreting between-arm contrasts.

\subsection{Engagement intensity}
``Engagement'' can conflate four distinct constructs, and the analysis keeps them separate: \emph{usage} (whether a student attends at all, and weekly completion), \emph{engaged time} (metered on-task minutes), \emph{interaction intensity} (turns, questions, and words produced per engaged minute), and \emph{attention} (paradata on where focus goes during a session). The paper's headline process contrast concerns interaction intensity on equal engaged time; usage and attention are reported separately in Section~4.2 and do not simply move in voice's favor. For the dose-response models, engagement intensity is measured from platform logs as (i) active tutor time on the week's focal topic (idle gaps excluded), with robustness co-measures of (ii) number of dialogue turns and (iii) number of on-topic retrieval/answer attempts. The two measures used throughout Sections~4--5 are computed as follows.

\noindent\textbf{Engaged time.} Engaged seconds accrue from platform telemetry under an idle clamp: a session's clock advances only within five minutes of the student's most recent input (utterance, keystroke, or blackboard interaction), with server-side caps, and is summed to the student-week. This is the measure that gates the 20-minute weekly goal, defines week completion, and enters the H4 dose-response models (as weekly minutes and as study-total hours). Because completion is gated on it, engaged time is variance-compressed near the goal by design, which we account for when interpreting time-on-tutor coefficients.

\noindent\textbf{Turn taking.} Turn density is student turns per conversation minute: the number of student turns in a session's transcript divided by the session's duration in minutes, aggregated to the student-week as total student turns over total session minutes, and to the student-modality cell as the mean over that student's weeks in the modality. Turn counts and durations come from the voice provider's conversation archive, re-pulled at analysis time: 291 archived conversations, the full archive including post-study usage, of which the 256 field-period study conversations of Table~\ref{tab:summary} are the subset. The archive covers 124 of the 145 analysis student-weeks (29 analysis-sample AI students $\times$ 5 weeks). Questions asked, response latencies, and utterance lengths in Table~\ref{tab:process} are computed from the same per-turn transcript records.

\noindent\textbf{Turn-level annotation.} To separate substance from conversational management, an LLM classifier also labeled every student turn (shown with the immediately preceding tutor turn as context) as \emph{effective} (a substantive answer, attempted answer, reasoning step, or genuine question about the material, regardless of length) or \emph{filler} (bare acknowledgements, ``let me think,'' pauses, greetings, and repeat requests). Questions were typed as \emph{substantive} (seeking new understanding) versus \emph{clarification} (asking the tutor to repeat or rephrase what it just said). Labels are context-dependent by design: a bare ``yes'' answering a direct content question is effective; the same word as acknowledgement is filler. These annotations support the descriptive interaction-pattern results of Section~4.2; we note a human-agreement validation of the classifier as future work (Section~\ref{sec:future}).

\subsection{Empirical Strategy (pre-registered)}
The analysis follows the study's two randomized layers. We pre-specified all confirmatory and secondary analyses before collecting outcome data.

\subsubsection{Notation}
Index students by $i$, weeks by $w\in\{8,\dots,12\}$, and assessment items by $k$. Let $AI_i\in\{0,1\}$ indicate assignment to AI access (1) versus holdout (0). Let $Voice_{iw}\in\{0,1\}$ indicate whether week $w$ is delivered in voice modality for AI-assigned student $i$ (text otherwise). Let $Pre_i$ and $Post_i$ denote total pre- and post-test scores.

\subsubsection{Primary analysis: AI access (H1)}
The primary estimand is the intent-to-treat effect of AI access on post-test total score, controlling for baseline performance (ANCOVA):
\begin{equation}
Post_i = \beta_0 + \beta_1\,AI_i + \beta_2\,Pre_i + \varepsilon_i.
\end{equation}
The pre-registered specification will include the stratification controls used in randomization, and inference will use robust standard errors.

\subsubsection{Secondary analysis: modality contrast (H2)}
\label{sec:h2spec}
Among AI-assigned students, we estimate the voice-vs-text contrast from within-student weekly alternation:
\begin{equation}
Y_{iw} = \alpha_i + \gamma_w + \delta\,Voice_{iw} + \varepsilon_{iw},
\end{equation}
where $Y_{iw}$ denotes week-level performance measures and $\alpha_i$ and $\gamma_w$ are student and week fixed effects. Standard errors will be clustered at the student level. Identification rests on the assessment's week-to-item mapping: each course week has exactly three post-test items (two MCQ at 3 points each plus one SOLO-scored short answer, 11 points per week), so $Y_{iw}$ for the learning outcome is a three-item subscore, and the modality effect is identified from within-student differences between voice-week and text-week subscores. Three-item subscores are noisy measures, and the design offsets that noise with scale (145 student-weeks with all between-student variance removed); we report equivalence bounds alongside the point estimate, so that a null can be read as affirmative evidence of equivalence rather than as a lack of power. $Voice_{iw}$ is the \emph{realized} modality from session logs (the dominant channel of that week's sessions), observed directly for 138 of the 160 weeks of the 32 post-test-completing AI students; for unattended weeks, the deployed alternation rule (calendar parity in weeks 1--2, then the opposite of the student's most recently completed week) supplies the value.

\subsubsection{Secondary analyses: Bloom interaction, dose-response, and depth (H3--H5)}
H3 is tested at the item level. With $g_{ik}$ the pre-to-post gain of student $i$ on item $k$ (MCQ items as 0/1 correctness differences; short-answer items as SOLO-level differences rescaled to unit range) and $High_k$ indicating higher-order Bloom items,
\begin{equation}
g_{ik} = \lambda_k + \theta_1\,AI_i + \theta_2\,(AI_i \times High_k) + \varepsilon_{ik},
\end{equation}
with item fixed effects $\lambda_k$ absorbing the Bloom main effect and standard errors clustered by student; $\theta_2$ is the H3 estimand. H4 relates learning gains to the pre-specified engagement measures (active topic time, turns, retrieval attempts) among AI-arm students; a two-stage least squares complier-average effect, with assignment as the instrument for engagement, complements this analysis. H5 examines SOLO-based depth outcomes (mean SOLO and share SOLO $\geq 3$) in specifications analogous to H1.

\subsubsection{Multiplicity and reporting}
Per the plan, we evaluate the primary test (H1) and the preregistered depth co-primaries (mean SOLO and share SOLO $\geq 3$) at $\alpha=0.05$ (two-tailed) without correction. The remaining secondary family (H2--H4 and related item-level checks) is controlled using Benjamini--Hochberg FDR at $q=0.05$. All estimates will be reported with 95\% confidence intervals and standardized effect sizes.

\section{Results}

\subsection{Implementation and participation}
The five-week field period (7/1--8/5/2026) finished on schedule, and protocol fidelity was high: no holdout student accessed the platform, and the 298 tutor sessions took place in their platform-assigned weekly modality. Weekly completion of the 20-minute engaged-time goal was stable in the AI arm, at 29/31/28/30/26 of 52 across the five weeks (50--60\%); holdout check-ins were 20/23/21/22/19 of 34 (56--68\%). At the student level, 27 AI students (52\%) and 20 holdout students (59\%) completed at least four of the five weeks, and 18 and 14, respectively, completed all five. Attrition was concentrated at the start: 14 AI students who consented never held a single session, and 7 holdout students never completed a check-in. This early non-use is a substantive finding about adoption rather than a mere nuisance: a quarter of the students offered a free tutor never tried it once. A tutor can therefore serve its users extremely well and still face a first-session adoption problem. Deployment design (defaults, course integration, first-week onboarding) must solve that problem, which makes it a lever for the planned deployment of Section~\ref{sec:future}. Post-test completion was 51 of 86 (59\%; 32 AI / 19 holdout, 62\% vs.\ 56\%, difference n.s.), statistically balanced across arms and higher than the study's own week-five participation; completers are also balanced on pre-test scores ($p=.68$). Because the outcome is observed only for post-test completers, the between-arm estimates are complete-case randomized comparisons. Balanced completion and baseline scores are reassuring but do not, by themselves, remove the selection concern. In post-hoc checks, pre-test score does not predict analysis inclusion differently across arms (logit interaction $p=.71$), and re-estimating the primary specification with inverse-probability weights (inclusion modeled on arm and pre-score) gives $+6.52$ ($p=.005$). Table~\ref{tab:summary} summarizes the study corpus.

\begin{table}[t]
\centering
\caption{Study data summary (field period 7/1--8/5/2026; post-study usage excluded).}
\label{tab:summary}
{\small
\setlength{\tabcolsep}{4pt}
\begin{tabular}{lr}
\hline
Participants (consented, randomized) & 86 \; (52 AI / 34 holdout) \\
AI start modality & 26 voice-start / 26 text-start \\
Weekly completions, AI arm & 29 / 31 / 28 / 30 / 26 of 52 \\
Weekly check-ins, holdout arm & 20 / 23 / 21 / 22 / 19 of 34 \\
Completed $\geq$ 4 of 5 weeks & 27 AI / 20 holdout \\
Tutor conversations & 256 \; (298 session rows) \\
Metered voice minutes & 1{,}880 \\
Student messages & 4{,}231 \; (2{,}526 voice / 1{,}705 text) \\
AI-arm weekly survey responses & 97 \; (32 students) \\
Instrumented typed turns (wks 11--12) & 621 \\
\hline
\end{tabular}}
\end{table}

\subsection{Process results: from composition to conversation}
\begin{table}[!tp]
\centering
\caption{Within-student process contrasts, weeks 8--12 (paired students with completed weeks in both modalities; $n=32$ for behavioral, $n=34$ for language measures; exact paired sign tests).}
\label{tab:process}
{\small
\setlength{\tabcolsep}{3.5pt}
\begin{tabular}{lcccc}
\hline
Measure & Voice & Text & Favoring & Sign $p$ \\
\hline
Engaged minutes per completed week & 27.5 & 28.0 & 14 v / 17 t / 1 tie & .72 \\
Student turns per minute & 1.34 & 0.75 & 31 of 32 voice & $<.001$ \\
Questions asked per student & 11.3 & 4.6 & 25 of 32 voice & .002 \\
Words per student message & 23.7 & 16.1 & 28 of 34 voice & $<.001$ \\
Median student reply gap (raw) & 27\,s & 54\,s & 26 of 27 faster in voice & $<.001$ \\
Sessions per completed week & 1.80 & 1.97 & --- & --- \\
\hline
\end{tabular}}
\end{table}

\begin{figure}[!tp]
  \centering
  \includegraphics[width=.92\textwidth]{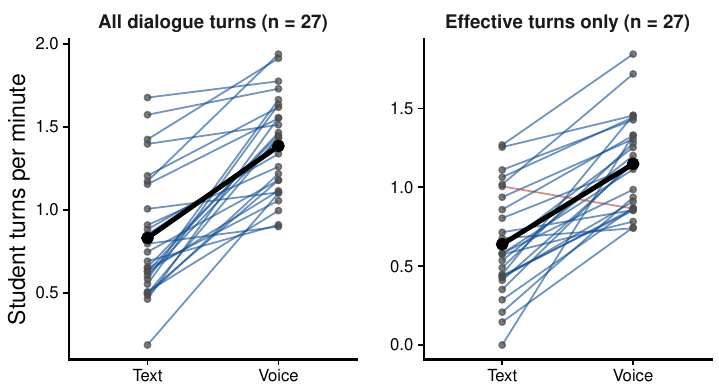}
  \caption{The process transformation, student by student: mean student turns per minute in text versus voice weeks, one line per paired student (blue: higher under voice; red: lower; black: arm mean). Left: all dialogue turns from the conversation archive (27 of 27 higher under voice). Right: classifier-labeled \emph{effective} turns only (26 of 27); the density premium consists of substantive turns rather than filler.}
  \label{fig:process}
\end{figure}

The within-student contrasts among the 32 students with completed weeks in both modalities are the study's first substantive result (Table~\ref{tab:process}). They describe a change in the \emph{kind} of interaction more than in its amount. With text, students deliberate and compose; with voice, they converse and think aloud, and they ask more questions. Engaged time is modality-neutral: paired means of 27.5 voice versus 28.0 text minutes per completed week (14 students higher in voice, 17 in text, 1 tie), exactly what the common 20-minute gate should produce. What students do with that time differs. In voice weeks the same students produce 1.34 dialogue turns per minute versus 0.75 in text weeks (31 of 32 students higher in voice) and ask 11.3 questions per student versus 4.6 (2.4$\times$; 25 of 32 higher in voice); question-asking is the behavior most tightly linked to learning in the self-explanation literature. The language register shifts as well. Spoken turns run 23.7 words versus 16.1 typed (28 of 34 paired students), carry hedges at 3.0$\times$ the typed rate (thinking aloud), and more often contain a question (15.5\% versus 11.7\% of messages), while length-controlled lexical diversity is essentially tied (TTR-100 0.66 versus 0.69), so the difference lies in the amount of language students produce rather than in the richness of their vocabulary. Because several of these measures are skewed ratios, we test each student's within-person direction with exact paired sign tests, which assume neither normality nor equal variances; with sign consistency this high (from 25 of 32 students on question-asking to 31 of 32 on turn density), the choice of distributional assumptions does not matter. An independent recomputation of turn density from the provider's conversation archive (Section~4.5) corroborates the headline contrast: 1.38 vs.\ 0.83 turns/min, higher under voice for 27 of 27 paired students (Wilcoxon $p<10^{-7}$). Figure~\ref{fig:process} shows the contrast student by student. On the archive measure every paired student converses more densely under voice, and restricting the count to classifier-labeled effective turns changes the picture for exactly one student.

Raw turn counts, however, treat ``let me think'' and a worked NPV attempt as equal, so the turn-level annotation of Section~3.9 separates them. The extra density is substantive: \emph{effective} turns per minute nearly double under voice (1.15 versus 0.64; higher under voice for 26 of 27 paired students, Wilcoxon $p<10^{-7}$), while the \emph{filler share} of turns is statistically indistinguishable across modalities (17\% versus 16\%, $p=.26$). In proportion, spoken interaction is as substantive as typed interaction; there is simply much more of it. Question-asking rises on both types: substantive questions run 4.1 versus 1.7 per hour ($p=.003$) and clarification requests 4.4 versus 1.8 ($p=.002$). The ephemeral audio channel does provoke more ``say that again,'' but it provokes just as much additional genuine inquiry. Effective utterances are also \emph{longer} under voice (25.4 versus 17.9 words per turn, $p=.002$): students think aloud in sentences they would never type. Together these annotations sharpen the headline result. On equal engaged time, voice substantially increased the number of substantive turns and genuine questions and the amount of language produced, while learning gains were preserved (Section~4.5).

\begin{figure}[!htbp]
  \centering
  \includegraphics[width=\textwidth]{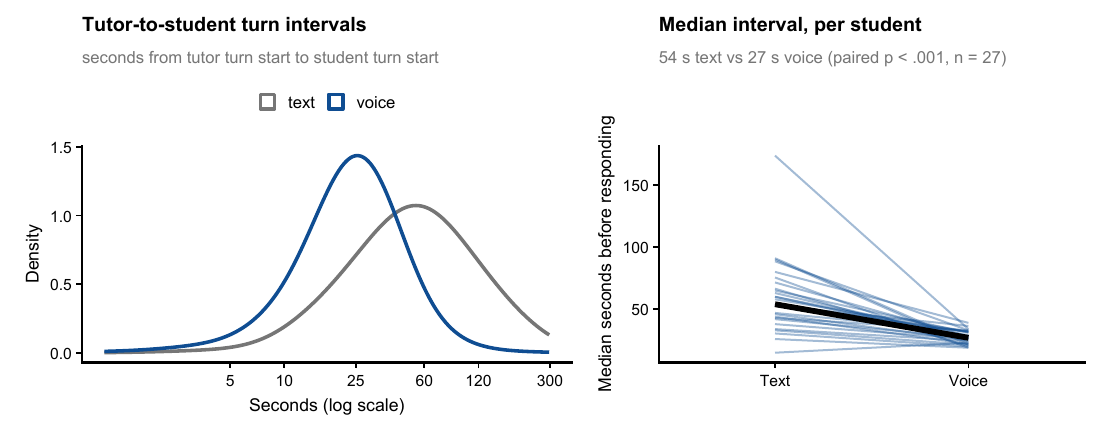}
  \caption{The thinking pause collapses under voice. Left: the distribution of intervals from the tutor's turn
  to the student's, pooled over 3{,}426 timed exchanges (log scale); the two channels barely overlap. Right:
  each student's own median interval in text and in voice, with the group median in black; 26 of 27 students
  reply faster in voice. The interval bundles listening or reading with thinking and, in text, typing, so it is
  not pure deliberation. Keystroke telemetry, however, decomposes the text gap into 25.9\,s before the first
  keystroke plus 13.9\,s composing, and subtracting each tutor turn's own playback time (17.5 characters/s) puts the median voice reply about 4\,s after the tutor stops.}
  \label{fig:tempo}
\end{figure}

Turn-taking is where the two channels come apart most sharply (Figure~\ref{fig:tempo}). Across 3{,}426 timed exchanges, the interval from the tutor's turn to the student's runs a median 25\,s in voice against 53\,s in text, and the same gap holds within student: comparing each student's own median across the two channels gives 27\,s in voice against 54\,s in text, with 26 of the 27 students who contribute paired latency data replying faster in voice (paired $p<.001$). The distributions barely overlap. Netting out audio playback turn by turn (a measured 17.5 characters/s applied to each tutor turn's own length) locates where that time goes: the median playback-adjusted gap is 3.6\,s, so the typical voice turn begins about 4\,s after the tutor stops speaking, and 12\% of turns begin \emph{before} it stops. Keystroke telemetry over the final two weeks decomposes the text gap directly: across 621 typed turns, a median 25.9\,s elapses before the first keystroke (22.1\,s of it attention-on after subtracting away-time) plus 13.9\,s composing. The text gap is deliberation plus typing rather than slowness; the voice gap leaves room for neither. Attention paradata qualifies the picture. Voice sessions tolerated more mid-session attention excursions in week~11 (away-share mean 26\% versus 14\%), and the two modalities converged by week~12 (18\% versus 19\%), while the engaged-time ledger's idle clamp binds almost exclusively in text (ledger-to-wall-clock 0.98 voice versus 0.66 text; text tabs sit open while students multitask). Modalities also absorb workload differently. When the final week handed students text they completed it 17 of 18 times versus 9 of 12 when handed voice, and text weeks split into more, shorter sittings (1.97 versus 1.80 sessions per completed week), consistent with text's flexibility under deadline pressure and voice's demand for a quiet, contiguous block. The construct distinctions of Section~3.9 earn their keep here. Voice raises interaction per engaged minute; it does not raise every engagement construct at once. Attention excursions were initially higher under voice, and when workload peaked students completed text weeks more readily. Voice concentrates production into denser exchanges, whereas usage and attention respond to modality in their own ways.

\subsection{Stated preference and experience}
Weekly surveys (97 AI responses from 32 students) show preference moving toward voice: mean preference on the $-2$ (text only) to $+2$ (voice only) scale moved $-0.21 \rightarrow +0.05 \rightarrow +0.11 \rightarrow +0.69 \rightarrow +0.29$ across the five weeks, and only the first week leaned toward text. Among 27 repeat responders, 12 drifted toward voice (average climb $+1.8$ scale points), 10 held position, and 5 drifted toward text; every downward drift stopped at ``mostly text,'' and voice sessions out-rated text sessions on quality in all five weeks (week 12: 4.83 versus 4.30 of 5). The share reporting increased confidence in the module's material rose monotonically from 54\% to 94\%, 94\% ended the study likely to continue using the tutor afterward, and the modal trust posture was calibrated: 67\% ``generally trust but verify,'' 15\% trusting without verification.

\subsection{The realized control condition: ambient consumer AI did not appear to close the gap}
The holdout's weekly check-ins confirm the preregistered expectation that the control condition is self-study \emph{with} ambient AI: 21--42\% used consumer tools for coursework in any given week (Claude, Gemini, and ChatGPT most often), so the randomized comparison evaluates the tutor against a realistic self-study environment in which consumer AI remained freely available. H1 therefore estimates the tutor's marginal value over that environment rather than over an AI-free baseline. The within-holdout split is observational, but it bears directly on the paper's motivating question. Among the 17 analysis-sample holdout students, the nine who reported external-AI use (one to five weeks of it) gained no more on the total score than the eight who reported none ($+10.3$ versus $+12.8$ of 55; a descriptive contrast on self-selected subgroups, $p=.45$), and both trailed the AI arm's $+15.1$. The preregistered per-protocol analyses provide a sensitivity check. Progressively \emph{removing} ambient-AI users from the control group should raise the tutor's estimated advantage if their AI use had been closing the gap; instead it leaves the estimate essentially unchanged (Figure~\ref{fig:forest}). Whatever holdout students did with consumer AI, it did not appear to close the structured-tutor gap, a pattern consistent with the offloading evidence that motivates the design \citep{bastani2025,darvishi2024agency}. Holdout weekly experience ratings declined over the module (3.78 to 3.21 of 5) as workload compounded, and open responses repeatedly named the support gap the tutor targets: on-demand explanation between office hours.

\subsection{Learning outcomes}
The post-test window closed August 15 with $N=51$ completers, every one with a linkable pre-test. The preregistered analysis sample is $N=46$ (29 AI / 17 holdout) after the quality screens fixed in the plan: four administrations fell under the 8-minute completion-time floor, and the response-quality screen excluded one for absent short-answer effort (all five post-test short answers left effectively unanswered, council SOLO $\approx 0$, by a student with a perfect pre-test MCQ; retaining this respondent yields $\beta=+5.52$, $p=.052$ on the primary specification). The four-provider council of Section~3 scored the short answers from both administrations with condition masked (510 responses; cross-provider linear $\kappa=0.81$; an all-Anthropic sensitivity panel agrees in magnitude throughout). Table~\ref{tab:outcomes} collects every preregistered estimate; the paragraphs below walk through them in order.

\begin{table}[t]
\centering
\caption{Preregistered learning-outcome estimates (analysis $N=46$: 29 AI / 17 holdout). ANCOVA rows report $\beta_{AI}$ from $Post = \beta_0 + \beta_1 AI + \beta_2 Pre\,(+\,strata)$; H2 reports the within-student voice coefficient with student and week fixed effects, clustered by student. $^{***}p<.01$, $^{**}p<.05$, $^{*}p<.10$ (raw, two-sided); bold labels mark the preregistered confirmatory tier, evaluated uncorrected at $\alpha=.05$ per the plan; the secondary family is Benjamini--Hochberg-controlled at $q=.05$ and is read as estimation rather than confirmation.}
\label{tab:outcomes}
{\small
\setlength{\tabcolsep}{2.5pt}
\begin{tabular}{llccc}
\hline
Hypothesis & Outcome & Estimate & 95\% CI & $p$ \\
\hline
H1 & Total score (of 55), baseline & $+3.56^{*}$ & $[-0.4,\,+7.5]$ & .079 \\
\textbf{H1 (primary)} & Total score, + randomization strata & $+6.63^{***}$ & $[+1.95,\,+11.31]$ & .007 \\
H5 & MCQ component (of 30) & $+0.88$ & $[-1.5,\,+3.3]$ & .46 \\
H5 & Short answer (of 25) & $+2.86^{**}$ & $[+0.14,\,+5.6]$ & .040 \\
\textbf{Co-primary} & Depth: share SOLO $\geq 3$ & $+0.221^{**}$ & $[+0.02,\,+0.42]$ & .029 \\
\textbf{Co-primary} & Depth: mean SOLO level & $+0.391^{**}$ & $[+0.01,\,+0.77]$ & .046 \\
H2 & Week subscore, voice vs.\ text (of 11) & $-0.01$ & $[-0.62,\,+0.60]$ & .98 \\
H2 & Equivalence, TOST $\pm0.3$\,SD & \multicolumn{2}{c}{equivalent$^{**}$} & .018 \\
H3 & AI $\times$ higher-order Bloom & $+0.06$ & $[-0.10,\,+0.21]$ & .48 \\
H4 & Gain per completed week (AI arm) & $+2.00^{**}$ & $[+0.37,\,+3.6]$ & .018 \\
H4 & CACE per engaged week (2SLS) & $+0.87^{*}$ & $[-0.03,\,+1.8]$ & .064 \\
\hline
\end{tabular}}
\end{table}

\textbf{Both arms learned.} Pre-to-post gains are large and unambiguous within arm: $+15.1$ of 55 points in the AI arm and $+11.4$ in holdout (both $p<10^{-5}$; a raw gain gap of $+3.7$), the expected pattern given that both arms took the same course. The between-arm question is the increment.

\textbf{H1 (access): supported under the full preregistered specification.} The AI arm gains $+6.63$ additional points (95\% CI $[+1.95, +11.31]$, $p=.007$) in the design-matched ANCOVA with randomization-strata fixed effects, the model the preregistration named primary; the covariate-light specification gives $+3.56$ ($p=.079$), and we report both so that the sensitivity to specification is visible (gain-score $d=+0.49$). The result holds on the same-provider sensitivity panel ($+5.22$, $p=.044$), and we report it with and without the response-quality exclusion above. Post-test MCQ performance reached the ceiling (46\% of all students scored a perfect 30/30), which is itself informative: the cohort mastered the surface content, so the instrument discriminates mainly in the written-reasoning half, which is where the treatment effect appears.

\begin{figure}[!t]
  \centering
  \includegraphics[width=.92\textwidth]{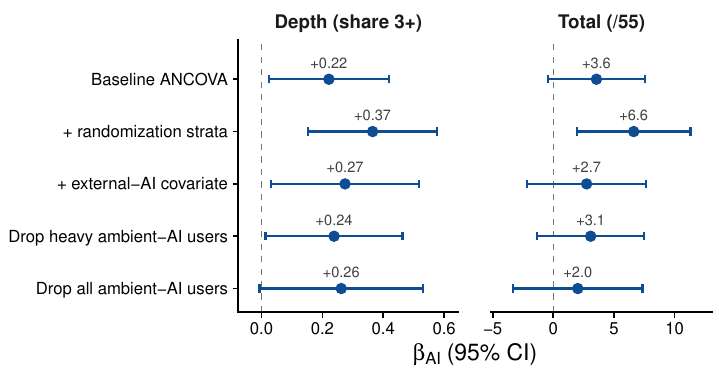}
  \caption{Specification stability of the access effect ($\beta_{AI}$, 95\% CI). Every specification points positive on both outcomes; the depth estimate barely moves.}
  \label{fig:forest}
\end{figure}

\begin{figure}[!t]
  \centering
  \includegraphics[width=.92\textwidth]{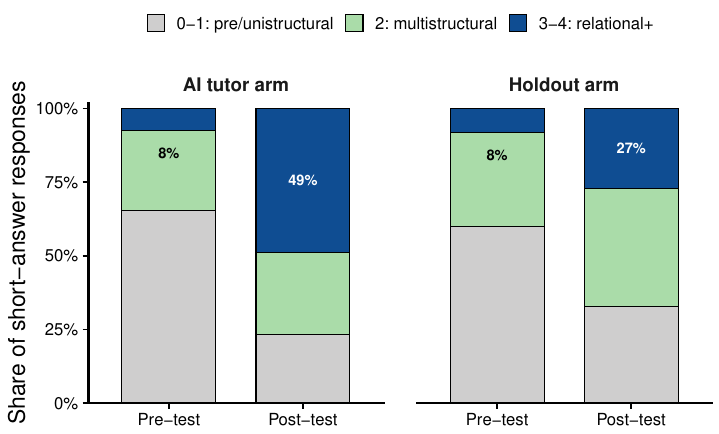}
  \caption{Depth of written answers by arm and administration: share of short-answer responses in each SOLO band. Both arms improve; the growth of relational-or-better answers (dark blue) is twice as large under the tutor (8\%$\to$49\% vs.\ 8\%$\to$27\%).}
  \label{fig:depth}
\end{figure}

\textbf{Depth over surface: where the signal lives.} The access effect concentrates exactly where the ceiling does not bind (Figure~\ref{fig:depth}). MCQ gains are indistinguishable between arms, while the share of a student's short answers at relational quality or better (SOLO $\geq 3$) grows sixfold under the tutor, from 8\% to 49\%, versus 8\% to 27\% in holdout (gain $d=+0.67$), and the AI arm leads on all five short-answer items. Both depth co-primaries clear conventional significance under the preregistered council (share: $\beta=+0.221$, 95\% CI $[+0.02, +0.42]$, $p=.029$; mean SOLO: $+0.391$, $p=.046$), as does the short-answer point total ($+2.86$, $p=.040$); the all-Anthropic sensitivity panel agrees in magnitude on the share outcome ($+0.200$, $p=.064$), and human adjudication of the flagged responses (Section~\ref{sec:future}) is the confirmatory follow-up. Figure~\ref{fig:forest} shows that this estimate is also the most specification-stable in the study: it moves between $+0.22$ and $+0.37$ across the strata, the external-AI covariate, and both per-protocol exclusions of ambient-AI-using holdout students.

\begin{figure}[!tp]
  \centering
  \includegraphics[width=.9\textwidth]{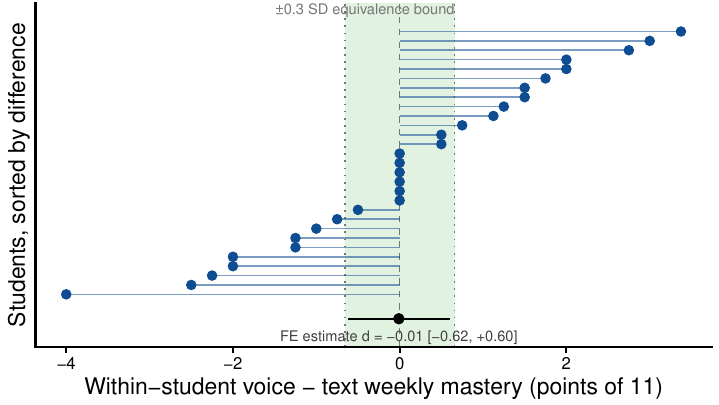}
  \caption{The modality equivalence (H2). Each blue point is one student's voice-minus-text difference in mean weekly mastery (29 students, sorted); the shaded band is the preregistered $\pm0.3$\,SD equivalence bound ($\pm0.66$ points of 11). Individual differences scatter widely in both directions, and the fixed-effects estimate with its 95\% CI (black, bottom) lies inside the bound: two one-sided tests reject any true modality difference larger than $\pm0.3$\,SD ($p=.018$).}
  \label{fig:h2eq}
\end{figure}

\textbf{H2 (modality): a precise equivalence.} Within students, measured week-level mastery (each week's three post-test items, 11 points) is statistically \emph{equivalent} between voice and text weeks over the five-week study: $\delta_{voice}=-0.01$ points (cluster $p=.98$; 145 student-weeks from 29 students), with two-one-sided-tests equivalence within $\pm0.3$\,SD ($p=.018$) and comfortably within $\pm0.5$\,SD ($p<.001$). We disclose one pipeline correction. An earlier build filled the unattended weeks by calendar parity throughout rather than by the deployed alternation rule of Section~\ref{sec:h2spec}, which mislabeled three unattended student-weeks; with those three weeks parity-coded the estimate is $-0.06$ ($[-0.65,\,+0.54]$, $p=.84$; TOST $p=.023$), so the correction changes nothing of substance. Figure~\ref{fig:h2eq} sets the estimate against the individual differences it summarizes: students scatter in both directions, and the fixed-effects estimate sits inside the preregistered bound.

 The equivalence extends to depth: the week's short-answer SOLO level shows no modality difference within student either ($\delta=+0.00$, $p=.99$). H2's directional prediction is not supported. The data instead support the stronger conclusion of statistical equivalence: voice neither outperforms nor underperforms text, and the process transformation of Table~\ref{tab:process} arrives with learning gains fully preserved.

\textbf{H3 (Bloom) and H4 (dose).} The Bloom interaction points in the predicted direction ($+0.06$, $p=.48$). Dose-response sharpens once the quality screens are applied: within the AI arm, each additional completed tutoring week is associated with $+2.0$ points of gain ($p=.018$ raw; Table~\ref{tab:outcomes} notes the multiplicity control for the secondary family), and the preregistered 2SLS complier estimate is $+0.87$ per engaged week ($p=.064$). Gains track cadence more than duration: engaged \emph{minutes} predict nothing between or within students (the completion gate compresses time variance by design), while completed \emph{weeks} do, a pattern consistent with a larger role for repeated, spaced practice than for total exposure time.

\textbf{Process-to-learning mediation (exploratory).} With outcomes in hand, we can test the process transformation directly as a predictor of learning. We re-derived per-student turn density from the full-archive re-pull of Section~3.9 (291 conversations covering 124 of the 145 analysis student-weeks, 27 students), which independently confirms the manipulation: 1.38 versus 0.83 student turns per minute, higher under voice for 27 of 27 students (Wilcoxon $p<10^{-7}$; Figure~\ref{fig:process}). The mediation step, however, is null. Within student, week-level mastery does not rise with that week's turn density (fixed-effects estimate $-0.20$ points per turn-per-minute, cluster $p=.63$), and across students the size of an individual's density gap between voice and text does not predict the size of their mastery gap between voice and text ($r=-.16$, $p=.41$). The conversational transformation is real and near-universal; its learning consequence is not visible in five-week subscores. This is consistent with the equivalence result: modality's value operates on the engagement margin rather than on the week-level mastery margin (Section~6).

In sum, the pattern across hypotheses is coherent rather than mixed: every estimate favors access, the signal is concentrated in conceptual depth, where the instrument has headroom, and the modality contrast, the best-powered estimand in the study, resolves as a precise equivalence. These are the signatures one expects of a real, medium-sized effect measured against an instrument whose surface half saturated. The planned course-integrated replication is powered to convert every secondary estimate into a confirmatory test.

\subsection{Course records: the instructor's own instruments}
\label{sec:grades}
The instructor's gradebook supplies three outcomes the study did not design: a midterm due in study week~1 whose content (course weeks 1--7) the tutor never taught, five weekly open-book knowledge checks on the study weeks' material, and a take-home final after the study. All 86 randomized students have all three on file, as totals and per question (110 items, Bloom-tagged from the exam text), so these contrasts use the intact randomization and the attrition concern of Section~4.1 does not arise. Table~\ref{tab:grades} reports the estimates in points of 100 under the preregistered specifications (the outcome itself was not preregistered).

On the final, the tutor arm scores $+2.57$ points above holdout on the 86 (95\% CI $[+0.12, +5.01]$, $p=.040$; $d=0.53$ in cohort SD) and $+2.53$ on the 46-student analysis sample ($p=.18$). The instrument is ceilinged: two-thirds of the class scored the maximum, the least-saturated of the 22 items still has 91.8\% of the class at full credit, and 77\% of the tutor arm reached the maximum against 56\% of holdout (Fisher $p=.057$), so every final estimate is attenuated. The placebo comes out as it should: on the pre-treatment midterm the arms differ by $+1.32$ ($p=.42$), and on the weekly-check mean by $+0.91$ ($p=.45$).

At the item level, the H3 model of Section~3.10, estimated with item and stratum fixed effects and the pre-test (Table~\ref{tab:grades}), gives $\theta_2=+8.91$ points of item credit on the final (95\% CI $[+0.97, +16.86]$, $p=.028$; $\theta_1=+1.83$ on the twenty lower-order items, $p=.072$); the PAP-literal model with item fixed effects only returns the identical $\theta_2$ ($p=.027$) and $\theta_1=+1.89$ ($p=.089$). Within the PAP's secondary family (H2, H3, and the H4 analog) its Benjamini--Hochberg $q=.084$, so the preregistered test does not pass at $q=.05$. The caveat concerns the higher-order stratum: the final has exactly two higher-order items, and the interaction rests on one of them (q4, analyze: $+15.7$, $p=.019$; q7, evaluate: $+5.8$, $p=.14$), on which 6 of 34 holdout students lost credit against 1 of 52 in the AI arm while the never-enrolled class stood at 97\%. The same model on the midterm returns $\theta_2=-3.89$ ($p=.11$), so there was no pre-existing higher-order advantage. On the weekly checks the arm contrast is $+1.02$ ($p=.34$), flat across Bloom levels.

The checks also carry H2: within tutored students, voice minus text is $-0.47$ on weekly totals ($p=.72$; 255 student-weeks, 51 students, the one student with a protocol modality override excluded) and $-0.14$ on items (95\% CI $[-2.24, +1.97]$, $p=.90$), with the exploratory quantitative-minus-qualitative split at $+1.21$ ($p=.60$). Modality is the platform's assignment (calendar parity in weeks 1--2, then the opposite of the student's most recently completed week), so from week~3 it depends on the student's own completion history; the calendar-parity coding fixed at randomization gives $+0.41$ ($p=.68$). The sign changes between codings but the null result does not.

Against the 395 classmates who never enrolled (a descriptive comparison), volunteers were not stronger on the pre-treatment midterm (holdout $-0.6$, $p=.71$; tutor $+0.6$, $p=.45$), the midterm-adjusted final shows no Hawthorne lift for the untreated holdout ($-1.2$, $p=.33$), and the tutor arm is $+1.1$ above the class on the same terms ($p=.012$). Within the holdout, the weekly check-in survey separates students who reported using a consumer LLM in at least one study week (17) from those who did not (11): the users scored lower on the midterm, on four of five weekly checks, and on the final, by two to seven points of 100 after adjusting for the pre-test, but none of the gaps is significant at $n=28$, and the midterm gap predates the study, so the pattern mirrors the within-holdout gain split of Section~4.4 without testing it. The records reproduce the direction of H1 on an instrument the study did not design and on the intact randomization, but the final's ceiling caps what they can show, and the higher-order signal rests on two items.

\begin{table}[t]
\centering
\caption{Tutor access on the instructor's own records (points of 100; positive = tutor arm higher). Totals: ANCOVA on pre-test and randomization strata, HC1 errors. Items (22 final, 28 midterm, 60 check items): item and stratum fixed effects with the pre-test, student-clustered errors; the voice--text rows use student and item fixed effects within the AI arm, under the platform's modality rule and under calendar-parity coding. Stars as in Table~\ref{tab:outcomes}; the H3 row's Benjamini--Hochberg $q$ within the PAP's secondary family is $.084$.}
\label{tab:grades}
{\small
\setlength{\tabcolsep}{2.5pt}
\begin{tabular}{llccc}
\hline
Record & Contrast & Estimate & 95\% CI & $p$ \\
\hline
Final total & AI $-$ holdout (86) & $+2.57^{**}$ & $[+0.12,\,+5.01]$ & .040 \\
Final total & AI $-$ holdout (analysis 46) & $+2.53$ & $[-1.27,\,+6.34]$ & .18 \\
Midterm total & AI $-$ holdout (placebo) & $+1.32$ & $[-1.91,\,+4.55]$ & .42 \\
Check mean & AI $-$ holdout & $+0.91$ & $[-1.46,\,+3.27]$ & .45 \\
Final items & AI $\times$ higher-order, $\theta_2$ (H3) & $+8.91^{**}$ & $[+0.97,\,+16.86]$ & .028 \\
Final items & AI, lower-order, $\theta_1$ & $+1.83^{*}$ & $[-0.16,\,+3.83]$ & .072 \\
Midterm items & AI $\times$ higher-order (placebo) & $-3.89$ & $[-8.65,\,+0.86]$ & .11 \\
Check items & AI $-$ holdout & $+1.02$ & $[-1.06,\,+3.10]$ & .34 \\
Check items & Voice $-$ text, platform rule (H2) & $-0.14$ & $[-2.24,\,+1.97]$ & .90 \\
Check items & Voice $-$ text, parity coding & $+0.41$ & $[-1.53,\,+2.36]$ & .68 \\
\hline
\end{tabular}}
\end{table}

\subsection{Ethics and accessibility}
The study ran under Boston University IRB Protocol \#8506E, and participants gave informed consent before randomization. Consent specified exactly what the platform stores (session transcripts, voice-session audio, blackboard content, interaction telemetry, and assessment responses), together with retention, access controls, and third-party processing by the voice-infrastructure and model providers. Research data are keyed by opaque study codes throughout: no identifying field enters the survey platform, the item-assignment logic (Appendix~\ref{app:assignment}), or the grading pipeline, and the scoring council is masked to identity, arm, and administration by construction (Appendix~\ref{app:council}). Incentives were identical across arms and independent of performance, no course grade depended on the study, and the tutor was barred from graded-assessment content. As for accessibility, the weekly alternation guaranteed that no participant depended on audio alone (every multi-week student's practice included text weeks), and within voice sessions the persistent blackboard rendered every worked result as readable, exportable text. The prompt's repeat-verbatim and think-time rules (Appendix~\ref{app:prompt}) accommodated non-native listeners in an internationally diverse cohort.

\section{Voice AI Economics}
Whether voice tutoring is worth deploying depends on what its process advantages cost, so the study prices every conversation individually from the provider's own billing metadata rather than from assumed rates. The infrastructure bills on two meters: it charges voice conversations per connected minute (covering speech recognition, synthesis, and connection time) and text conversations per message; on the plan used here, a credit is worth \$0.0001 and the LLM's token charges are billed \emph{inside} each conversation's credit cost rather than as a separate line. An idle text tab therefore costs nothing, while an open voice line keeps metering, an asymmetry that bears on how a deployment should be designed.

Five weeks of tutoring for the 52-seat AI arm cost \$562.59 in total: \$413.46 voice and \$149.13 text, a 2.8$\times$ ratio on similar engaged time (Table~\ref{tab:pricing}). Unit costs were stable across the study at \$0.22 per voice minute (1,880 metered minutes) and \$0.0165 per text message (9,038 messages). LLM charges were 69.5\% of total spend (\$390.88), so even in voice the language model, rather than the audio stack, was the largest cost. Per capita, the study cost \$10.82 per enrolled seat, \$14.81 per active user (38 students held at least one session), and \$3.91 per completed student-week across the 144 completed weeks.

\begin{table}[t]
\centering
\caption{Voice versus text pricing, priced per conversation from provider billing metadata (256 conversations; credits at \$0.0001; LLM token charges billed inside conversation cost).}
\label{tab:pricing}
{\small
\setlength{\tabcolsep}{4pt}
\begin{tabular}{lcc}
\hline
 & Voice & Text \\
\hline
Billing basis & per minute & per message \\
Measured unit cost & \$0.220/min & \$0.0165/msg \\
Metered volume & 1{,}880 min & 9{,}038 msgs \\
Total spend (5 weeks) & \$413.46 & \$149.13 \\
Weekly spend (\$) & 64/94/91/103/62 & 28/28/37/25/31 \\
Idle-tab cost & meters continuously & none \\
\hline
Study total & \multicolumn{2}{c}{\$562.59 \; (LLM share \$390.88 = 69.5\%)} \\
Per seat / active user / completed wk & \multicolumn{2}{c}{\$10.82 / \$14.81 / \$3.91} \\
\hline
\end{tabular}}
\end{table}

Two implications follow. First, the learning-equivalence result puts a price on the modality decision. Voice does not buy more measured learning per week (Section~4.5), so its 2.8$\times$ premium currently buys interaction density, question-asking, preference, and adherence, forms of engagement whose downstream value a scaled deployment can measure. Programs that value those process outcomes (or the adoption they may drive) have a concrete price for them; programs that do not can deploy text at a third of the cost and, under this provider stack and observed usage, lose no detectable measured weekly mastery within the prespecified equivalence bound. Second, the cost structure argues for the hybrid configuration the platform now supports, in which a single session carries both channels and students switch freely: per-message text absorbs long stretches of reading and thinking at near-zero cost, and the voice channel is kept for the retrieval-and-explanation exchanges that motivate it.

\section{Discussion}

\subsection{Design implications}
Three implications follow for the design of AI tutors. First, \emph{the pedagogy is the treatment}: the anti-offloading results rest on a tutor that diagnoses before explaining, requires an attempt before feedback, and closes topics with transfer. That behavior is a configuration choice rather than a property of the model, and any deployment can adopt or omit it. The contrast with the exoskeleton result of \citet{wiles2024exoskeleton} makes the stakes concrete. There, generative AI raised what workers could produce while it was present and left their unaided knowledge unchanged. Here, every learning outcome was measured with the tutor absent, on our tests and on the instructor's, and the gains survived. What the tutor did while present was refuse to be an exoskeleton: it withheld answers, required an attempt, and closed each topic only on unassisted transfer. Second, \emph{modality is, on this evidence, a lever on engagement rather than on learning}: voice changes how students interact (conversation instead of composition, twice the turns, 2.4$\times$ the questions) while week-level mastery stays equivalent, so the modality decision should be priced on engagement, adherence, preference, and cost rather than on expected score gains. Third, \emph{adoption is a first-class design problem}: a quarter of the students offered a free tutor never tried it once, so defaults, onboarding, and course integration plausibly matter as much as tutoring quality. The equivalence result frees designers to route that adoption problem through whichever channel each student will actually use; a hybrid voice-plus-text session (which the platform now supports as a single continuous conversation) lets students give the expensive channel to retrieval-and-explanation exchanges and the cheap one to reading and reflection.

\subsection{Why voice did not out-teach text: speed, reflection, and the value of friction}
The modality equivalence has a candidate explanation: the thinking pause collapses under voice (Section~4.2, Figure~\ref{fig:tempo}). What remains interpretive is what that pause was doing. Composing text is slow by construction. The keystroke telemetry shows a median 26 seconds of deliberation before the first keystroke plus 14 seconds of composing per typed turn, and the annotation results show that typed turns carry no more filler than spoken ones, so the friction is explicit construction rather than waste. That construction may itself be encoding: writing an answer forces the student to hold the pieces and assemble them, which is the integrative act the SOLO rubric rewards. Voice removes that friction and applies its own pressure in reverse. A conversational cadence in which the median turn begins about four seconds after the tutor stops speaking, and one turn in eight begins before it stops, rewards fast response over reflection. One participant described the tension directly: she reported repeatedly asking the tutor to pause so she could think and read the blackboard content before answering. The mediation null fits this reading. Turns per minute measures how fast an exchange moves rather than how deeply the student thinks, and a fast, dense exchange can be shallow, which is why conversational density does not predict mastery gains (Section~4.5) even though it is the modality's most pronounced signature. On this account, what voice demonstrably delivers is \emph{lower interaction cost} rather than more learning per unit of time: students do not have to type or compose, they ask more questions, and their preference for the modality persists. Voice is an adoption technology rather than an encoding technology. Learning gains held while engagement roughly doubled, and the design question this raises for the next study is whether lower interaction cost converts into more sessions, longer retention, and, through accumulated dose (Section~4.5's cadence gradient), more learning in the end.

\subsection{What the instructor's records add}
The instructor's gradebook (Section~\ref{sec:grades}) answers three questions that our own instrument could not. Our primary estimate rests on the students who finished both of our tests. His final is on file for every randomized student, and on that intact sample tutor access still adds 2.6 points of 100 ($p=.040$), so the result does not depend on who stayed. Our test was written and graded by us. His was written and graded by him, with the tutor barred from its content, so the same direction there argues against test familiarity as the mechanism. His midterm covers material the tutor never taught and shows no gap between the arms, a placebo our design could not run on itself. The records also bound two concerns raised below: volunteers scored no higher than the 395 classmates who never enrolled, and the untreated holdout's final did not rise relative to that class. What the records cannot do is confirm the Bloom hypothesis. The tutor arm's edge on the final sits on its two higher-order items, the midterm placebo leans the other way, and the interaction fails the preregistered multiplicity rule, so we read it as consistent with H3 rather than as evidence for it. The final's ceiling also caps every estimate drawn from it: two-thirds of the class scored the maximum, so the 2.6 points is attenuated by construction.

\subsection{Limitations}
Scale bounds the between-arm comparison: with 86 consented students, H1 is powered for medium-to-large effects, and Section~4.5 reports the estimate across every analytic fork (specification, exclusion, scoring panel) so that readers can inspect how it behaves. The instructor's final, on file for all 86, shows the same direction without any attrition (Section~\ref{sec:grades}), though its ceiling limits what it can add. The design concentrates its power on the novel question: the within-student crossover observes the same 32 students in both modalities across five weeks, and the process contrasts of Table~\ref{tab:process} have high within-student sign consistency (25 of 32 students for question-asking, 31 of 32 for turn density, 28 of 34 for message length, and 26 of 27 for reply gap), which no plausible resampling overturns. Stratified assignment and ANCOVA on the pre-test further tighten the between-arm estimate. Generalization is deliberately narrow: one quantitative module in one online MBA, a setting whose analytical, multi-step content plausibly favors text's persistence, though we treat that as an untested moderator rather than an established handicap for voice; Section~\ref{sec:future} discusses the replication path. Finally, consent-based enrollment means participants are self-selected within the cohort, as in any opt-in field study of teaching tools, and participants knew they were being studied, so Hawthorne and demand effects may inflate engagement and self-reported outcomes among volunteers aware of observation. Three design features and one external check bound this risk. The primary outcomes are objective test scores rather than self-reports; the process measures are passive telemetry rather than solicited behavior; and the within-student modality contrast holds observation constant across conditions, so awareness of being studied cannot by itself manufacture a voice--text difference. The external check is the instructor's gradebook (Section~\ref{sec:grades}): on his pre-treatment midterm, volunteers scored no higher than the 395 classmates who never enrolled, and the untreated holdout's final did not rise relative to that class once the midterm is held constant. A course-integrated deployment (Section~\ref{sec:future}) would remove the mechanism at its source: with the tutor shipped as standard course infrastructure to an entire cohort rather than as an opt-in experimental artifact, usage and outcomes would be observed under ordinary course conditions rather than study conditions.

\subsection{Future research}
\label{sec:future}
Six extensions build on the present evidence. \emph{Human adjudication:} the 16\% of council-scored responses that carry cross-provider disagreement flags await the preregistered human-expert pass (council--human $\kappa$ reported); because the primary and depth co-primaries rest on council-scored short answers, that pass is the natural confirmatory complement to the two scoring panels, which already agree in magnitude throughout. \emph{Course-exam triangulation:} the instructor's final in Section~\ref{sec:grades} had almost no headroom, so it could confirm the direction of the main effect but little else; a final built to discriminate, with open responses scored on SOLO, would turn that check into a test of the depth result. \emph{Selection and Hawthorne bounds:} the descriptive contrasts against never-enrolled classmates in Section~\ref{sec:grades} are a first bound, and comparing consented-but-never-engaged students with non-participants would sharpen it. \emph{Classifier validation:} a human-labeled sample would validate the turn-annotation classifier. \emph{Modality and language:} the voice--text contrast for non-native speakers, for whom listening and speaking loads may shift the modality trade-off, is a preregistrable moderator at cohort scale. \emph{Assessment integrity:} conversational, AI-administered knowledge checks raise the question of whether oral assessment deters or enables academic dishonesty relative to static quizzes; we take this up in a companion design.

The natural next study is a \emph{course-integrated deployment at cohort scale}: the tutor and its assessments shipped as standard course components rather than as an opt-in study, in a subsequent term. Power arithmetic from the observed effects says what that buys. Roughly 110 analyzed completers would power the \emph{unstratified} H1 specification (the stratified version is already significant here), about 100 would power the depth outcomes under either scoring panel and the complier-average dose effect, while the Bloom interaction at its observed size remains exploratory below several hundred. Because opt-in completion rather than enrollment caps the analyzable sample, course-integrated assessment (which converts completion from a favor into a requirement) is worth more than any increase in enrollment. Beyond replication, the adoption reframing of Section~6.2 defines its own agenda: under free modality choice rather than assigned alternation, does voice's lower interaction cost convert into more sessions, longer retention, and larger accumulated dose? The hybrid voice-plus-text session the platform now supports (one continuous conversation, channel switched at will) instruments that question directly, since it logs every switch as a revealed preference.

\section{Conclusion}
This paper reports a completed preregistered randomized field experiment that tests whether structured conversational AI tutoring improves learning in a graduate online MBA course and whether voice versus text modality matters. The evidence establishes a clear and near-unanimous process difference. On identical engaged time, voice tutoring doubles interaction density, multiplies question-asking by 2.4$\times$, shifts language toward the spontaneous spoken register, and wins students' stated preference over repeated exposure, while measured week-level mastery stays statistically equivalent within the prespecified bound, at a measured 2.8$\times$ cost premium. Modality changes how students interact with an AI tutor far more than it changes what they measurably learn from it over five weeks. Access to the tutor improved learning: the data support the preregistered primary under its full specification ($+6.6$ of 55 points, $p=.007$, robust across two scoring panels), with the gain concentrated in the conceptual depth of written reasoning (where relational-quality answers grew sixfold) rather than in ceilinged multiple-choice scores; and the control arm's own ambient-AI users, who gained no more than abstainers, descriptively illustrate the difference between ambient access and structured practice. The instructor's own gradebook, on file for every randomized student, shows the same direction on his final and no difference on a midterm the tutor never covered, so the result rests neither on our instrument nor on who completed our tests. The next deployment is designed to test whether these effects replicate at cohort scale under ordinary course conditions. Scaled online programs weighing AI tutoring configurations can reuse the design, instrumentation, and cost accounting reported here, and the guardrailed-tutor framing gives a concrete answer to the offloading problem of unguided AI: AI that makes the student do the thinking, rather than less AI.

\begin{credits}
\subsubsection{\ackname}
We thank the leadership of Boston University Questrom School of Business's Online MBA (OMBA) program for supporting this research. We are especially grateful to Lisa Rohrer and Jay Zagorsky for their help and guidance throughout the study. We also thank our voice infrastructure provider, ElevenLabs, for study support.

\subsubsection{\discintname}
The authors have no competing interests to declare beyond the acknowledged infrastructure support from ElevenLabs, which had no role in study design, analysis, or reporting.
\end{credits}

\clearpage
\appendix
\section{Item-variant assignment routine}
\label{app:assignment}
The listing below is the core of the counterbalancing generator (TypeScript; the full script is in the study repository). A seeded PRNG makes every batch reproducible. The balanced scheme equalizes each item's variant exposure across the cohort while each participant's own vector stays random.

\begin{figure}[!htbp]
  \centering
  \includegraphics[width=.96\textwidth]{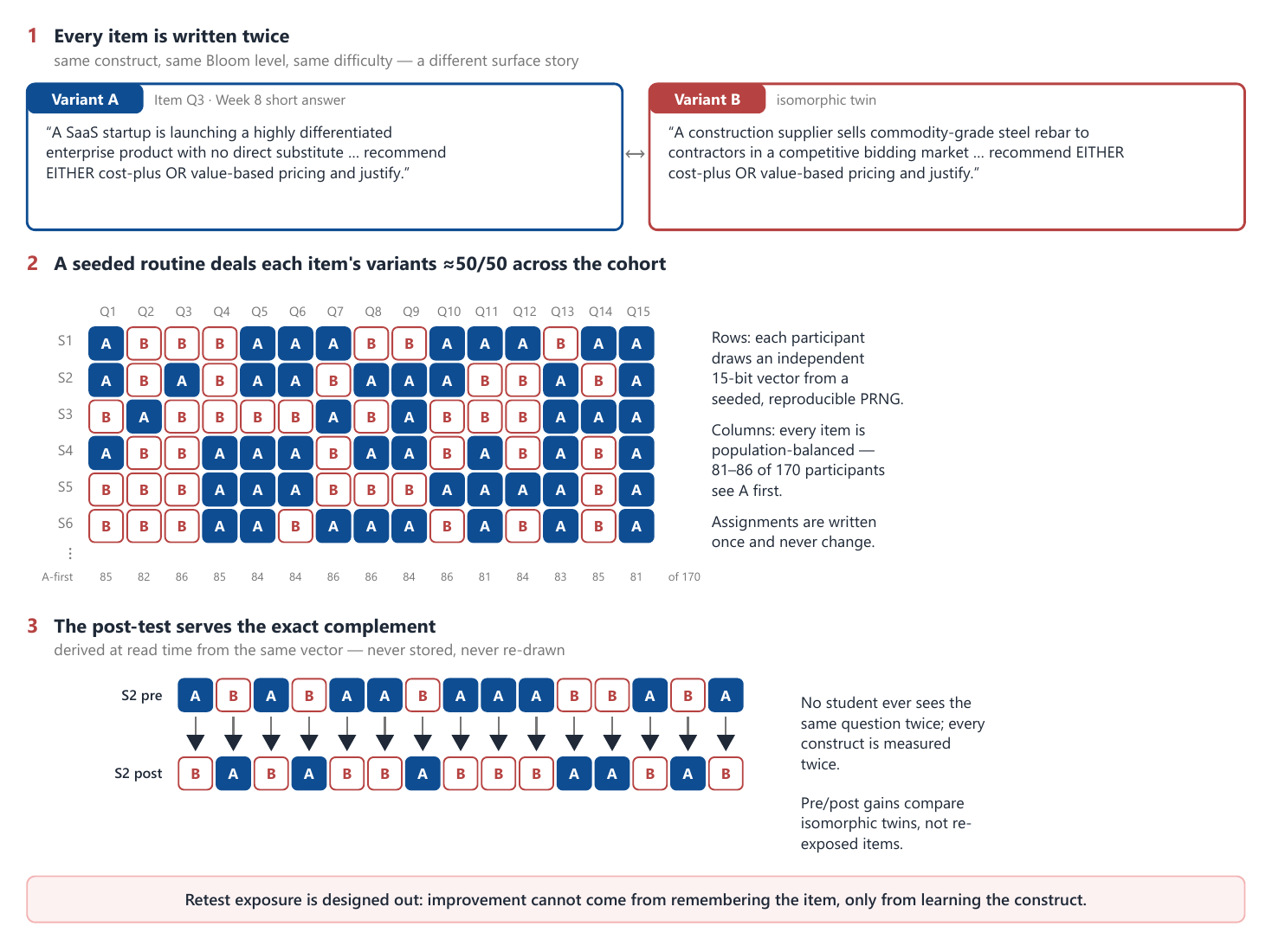}
  \caption{Isomorphic variant assignment, drawn from the deployed assignment table. (1)~Each of the 15
  items has two variants that share construct, Bloom level, and difficulty (the Week~8 short-answer
  pair is shown verbatim). (2)~Rows S1--S6 are the first six real assignment vectors (opaque study codes
  omitted); each column's variant exposure is balanced across the 170 assigned participants (81--86
  see A first). (3)~The post-test serves the exact complement of the stored pre-test vector, so no
  student meets the same question twice, and variant difficulty cancels out of within-student gains
  by construction.}
  \label{fig:variants}
\end{figure}

{\footnotesize
\begin{verbatim}
// Deterministic PRNG - a run reproduces exactly given SEED.
function mulberry32(seed: number): () => number {
  let a = seed >>> 0;
  return function () {
    a |= 0; a = (a + 0x6d2b79f5) | 0;
    let t = Math.imul(a ^ (a >>> 15), 1 | a);
    t = (t + Math.imul(t ^ (t >>> 7), 61 | t)) ^ t;
    return ((t ^ (t >>> 14)) >>> 0) / 4294967296;
  };
}

// 'balanced': for EACH item independently, build a column of n bits
// as close to 50/50 as possible, shuffle it, and deal it down the
// participants -- every item is population-balanced while each
// participant's vector stays random across items.
function generateBalanced(n, len, rng): number[][] {
  const vectors =
    Array.from({ length: n }, () => new Array(len).fill(0));
  for (let i = 0; i < len; i++) {
    const ones = Math.floor(n / 2) +
      (n % 2 === 1 && rng() < 0.5 ? 1 : 0);
    const col = new Array(n);
    for (let k = 0; k < n; k++) col[k] = k < ones ? 1 : 0;
    shuffleInPlace(col, rng);               // Fisher-Yates under rng
    for (let k = 0; k < n; k++) vectors[k][i] = col[k];
  }
  return vectors;
}

// PRE shows variant A where bit = 1 (else B); POST serves the
// complement, derived at read time and never stored. Unassigned
// participants are sorted by ID (deterministic), assigned in one
// batch, and rows are inserted plain (never upserted): existing
// assignments are immutable. study_code is an opaque Crockford
// base-32 token; no identifying field enters the vector logic.
\end{verbatim}
}

\section{Tutor system prompt (abridged)}
\label{app:prompt}
A structured system prompt governs the tutor's behavior (315 lines in the voice variant; a text variant differs only in channel policy and chat-panel behavior, and a post-study variant removes phase gating). The section structure below is verbatim. We condense the bodies but reproduce the interaction rules and the question ladder in full. Complete prompts are available from the authors.

\paragraph{Role.} ``You are the OMBA Module~2 Voice Tutor \dots You run spoken, oral practice sessions that help students rehearse and deepen their understanding of course material through structured Socratic questioning and constructive feedback. This is practice only: no grades are assigned. Your goal is not to test --- it is to walk each student up from recall to genuine transfer.''

\paragraph{Non-negotiable interaction rules (verbatim).}
\begin{enumerate}
\item \emph{One Question Rule.} Ask only one question per turn.
\item \emph{Groundedness.} Use only course material available through the knowledge bank (RAG) and the session context provided. Never guess the date, week, or topic --- they are provided.
\item \emph{Repeat Verbatim.} If asked to repeat a question, repeat it exactly.
\item \emph{Think Time.} After asking, allow silence ($\sim$30\,s) before offering to repeat.
\item \emph{No Long Spoken Lists.} Prefer single, open-ended prompts.
\item \emph{Scope and Time Gating.} What the student may practice depends on elapsed engaged time (below).
\end{enumerate}

\paragraph{Scope and time gating.} The app injects authoritative session context at connect (date, course week, topic, in-scope concepts, session clock, cross-session history). Until the week's 20-minute engaged requirement is met (Phase~1), practice is restricted to the current week's topic; afterwards (Phase~2), any course week is available: earlier weeks as review, upcoming weeks as clearly labeled previews that teach first, stay grounded in the knowledge bank, and never rehearse graded-assessment content.

\paragraph{Visual blackboard contract.} The tutor writes structured notes (Markdown with rendered LaTeX, tables, and chart specifications) beside the conversation through \texttt{display\_content} tool calls, on nearly every turn; it speaks naturally while the board renders, never reads LaTeX aloud, and keeps each write to one focused concept. A requested chart must be a real rendered chart, never a prose description.

\paragraph{Required flow.} (1)~An app-delivered opening adapts to the student (new; returning with requirement met; returning without) and asks whether they brought questions; student questions become attempt-first practice seeds. (2)~A focal topic is confirmed in one sentence. (3)~Questions climb Bloom's ladder in ascending order, one level at a time, each level mapped to a target SOLO level:

\begin{center}
\small
\begin{tabular}{clll}
\hline
Step & Bloom's level & Student task & Target SOLO level \\
\hline
1 & Remember & Recall the key term / definition & Unistructural \\
2 & Understand & Explain in own words; give the mechanism & Multistructural \\
3 & Apply & Use the concept in a familiar course case & Multi.\ $\rightarrow$ Relational \\
4 & Analyze & Compare; surface a tradeoff or pitfall & Relational \\
5 & Evaluate & Judge / justify a decision under conditions & Relational \\
6 & Create & Generalize to a NEW, unseen situation & Extended Abstract \\
\hline
\end{tabular}
\end{center}

\noindent Questions are generated only from retrieved course material, never repeated across a student's sessions (cross-session history lists prior questions), and the Create-level transfer question is the capstone of every topic. (4)~Each question follows ask $\rightarrow$ wait $\rightarrow$ structured feedback (verdict; what went well; what to improve; a structural suggestion; one micro-hint), with adaptive difficulty in both directions and a once-per-session, kindly delivered nudge when a clear pattern of superficial engagement appears (e.g., repeated summary requests with no attempts), logged silently for research. (5)~A topic is complete only after the student attempts a genuine transfer question (SOLO Extended Abstract), scaffolded until \emph{they} produce the generalization. (6)~Wrap-up summarizes strengths, review targets, and next steps.

\paragraph{Guardrails.} No full solutions to graded homework or exams; no help gaming assessments; preview without spoiling; redirect drift gently.

\section{LLM grading-council prompts}
\label{app:council}
Every council member (Section~3.7) received the same system prompt and per-item message. We key responses by opaque grade identifiers and shuffle them, so no model observes student identity, arm, or administration. The system prompt and the frozen rubric appear verbatim below.

\paragraph{System prompt (verbatim).}
{\footnotesize
\begin{verbatim}
You are one member of an independent grading council scoring graduate
business students' short-answer exam responses on the SOLO taxonomy.
You never know which student, group, or test administration a response
came from -- score each response on its own merits against the rubric
and the item's instructor anchor. Return STRICT JSON only: an array of
objects {"gid": string, "solo": integer 0-4, "why": one sentence
grounded in the rubric}. No markdown fences, no commentary.
\end{verbatim}
}

\paragraph{Frozen rubric (verbatim).}
{\footnotesize
\begin{verbatim}
SOLO taxonomy (Biggs & Collis, 1982) -- frozen scoring rubric:
  0  Prestructural   -- irrelevant, blank, or no valid engagement
                        with the task.
  1  Unistructural   -- exactly one relevant point; misses the
                        broader structure.
  2  Multistructural -- several relevant points, but listed without
                        integration.
  3  Relational      -- points integrated into a coherent, justified
                        argument that answers the question asked.
  4  Extended abstract -- integrates AND generalizes beyond the prompt
                        (transfers the principle, notes boundary
                        conditions, or connects to a broader framework).
Score the STRUCTURE of the response, not its length or eloquence. A
response that asserts the right conclusion with no reasoning is at most
Unistructural. Reasoning that is partially wrong can still be
Multistructural if several relevant considerations are present; factual
errors that undermine the argument cap the score at 2. Reserve 4 for
genuine generalization beyond what the prompt requires.
\end{verbatim}
}

\paragraph{Per-item message.} Each batch message contains the rubric above; the deployed question text of every variant present in the batch, each paired with its instructor anchor for a strong (Relational-or-better) answer; and up to twelve student responses labeled only by grade identifier and variant. The model returns one integer SOLO score and one rubric-grounded sentence per response; batches with missing or malformed entries are re-queried per response.

\section{Progress and monitoring interfaces}
\label{app:interfaces}
\begin{figure}[!htbp]
  \centering
  \includegraphics[width=.88\textwidth,page=1]{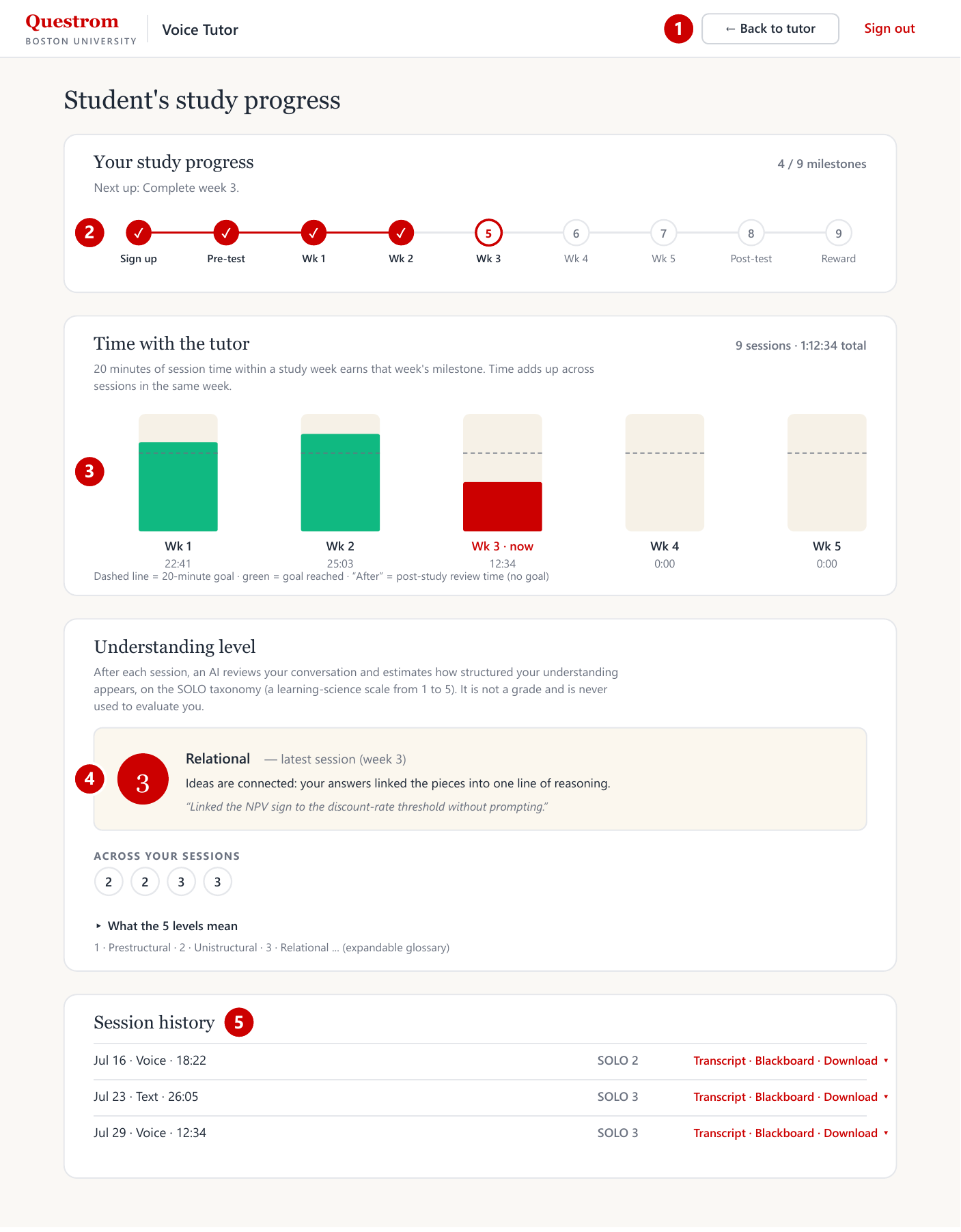}
  \caption{The student progress page (\emph{My progress}), redrawn from the deployed interface with representative values. (1)~Return-to-tutor navigation. (2)~The nine-milestone study tracker, sign-up through reward. (3)~Weekly tutoring-time bars against the dashed 20-minute goal line: green marks earned weeks; the current week accrues in scarlet. (4)~Session-level SOLO \emph{understanding level} feedback (an AI estimate of how structured the student's reasoning appears, explicitly framed as reflective and not a grade), with a cross-session history and an expandable glossary of the levels. (5)~Session history with per-session SOLO and transcript, blackboard, and download access.}
  \label{fig:studash}
\end{figure}

\begin{figure}[p]
  \centering
  \includegraphics[width=.78\textwidth,page=1]{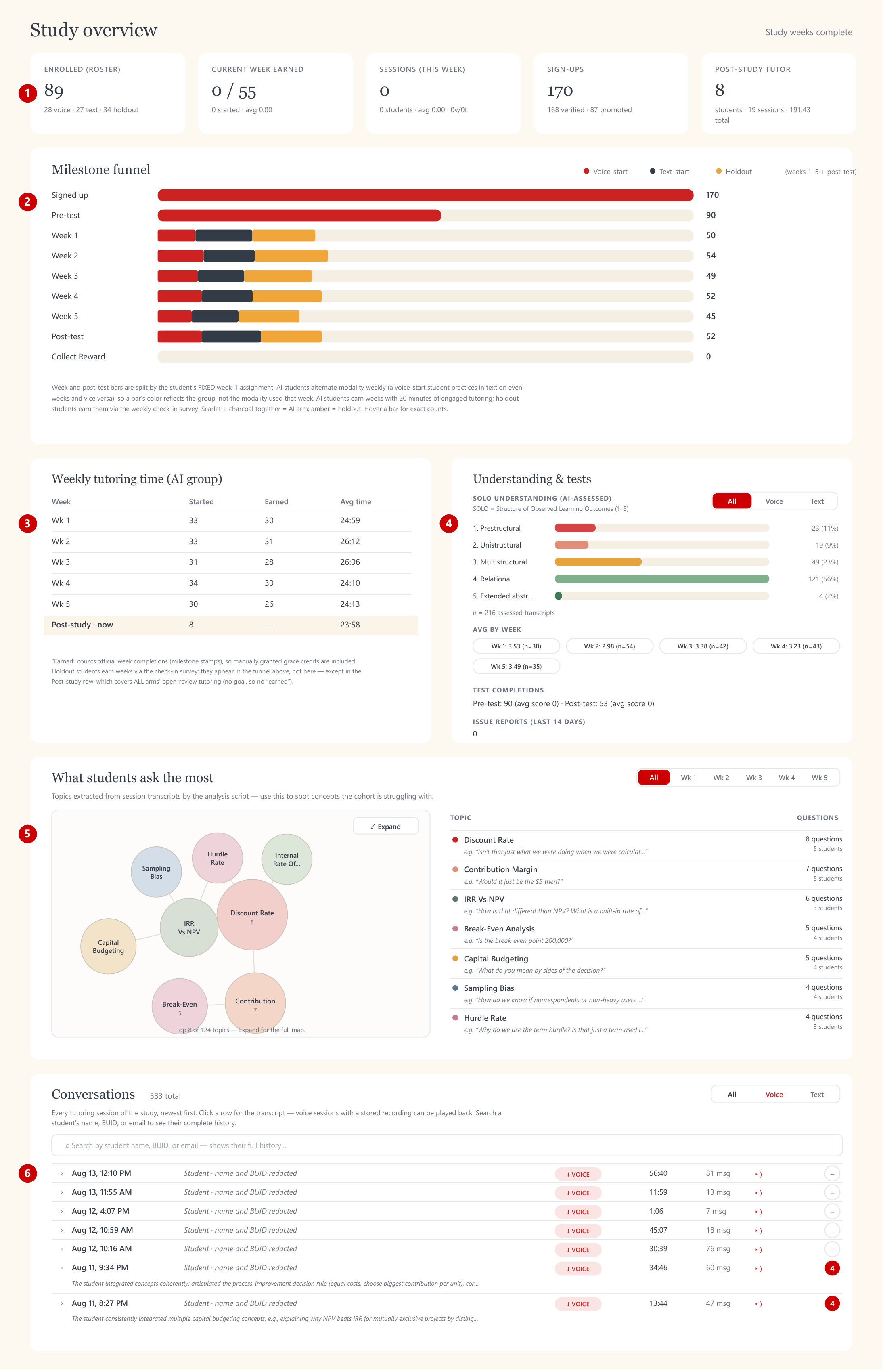}
  \caption{The instructor dashboard, redrawn from the deployed view at study close with its exact values; participant names, IDs, and identifying summary details are redacted. (1)~Roster and activity cards. (2)~Milestone funnel split by fixed week-1 assignment (scarlet $+$ charcoal $=$ AI arm; amber $=$ holdout). (3)~Weekly tutoring-time table for the AI group, with the post-study open-review row. (4)~AI-assessed session SOLO distribution ($n=216$ transcripts), weekly averages, and test completions. (5)~Topic bubbles and ranked question clusters from session transcripts (top 8 of 124). (6)~Conversation browser over all 333 sessions: transcripts, stored voice audio, AI summaries, and session-level SOLO badges.}
  \label{fig:instrdash}
\end{figure}

This appendix adds two further views of the platform: the student-facing progress page behind the header's \emph{My progress} control (Figure~\ref{fig:studash}) and the instructor dashboard used for study monitoring (Figure~\ref{fig:instrdash}). Both are redrawn from the deployed interface. The student page shows representative values; the instructor dashboard reproduces the deployed view at study close with its exact values, and participant identities are redacted.

\clearpage
\bibliographystyle{splncs04}
\bibliography{references}
\end{document}